# Patient-specific computational forecasting of prostate cancer growth during active surveillance using an imaging-informed biomechanistic model

G. Lorenzo[1,2], Jon S. Heiselman[3,4], Michael A. Liss[5], Michael I. Miga[3,6,7], Hector Gomez[8], Thomas E. Yankeelov[2,9,10], Alessandro Reali[1], Thomas J. R. Hughes[2]

[1]Department of Civil Engineering and Architecture, University of Pavia, Italy

[2]Oden Institute for Computational Engineering and Sciences, The University of Texas at Austin, USA

[3]Department of Biomedical Engineering, Vanderbilt University, USA

[4]Department of Surgery, Memorial Sloan-Kettering Cancer Center, USA

[5]Department of Urology, The University of Texas Health Science Center at San Antonio, USA

[6]Vanderbilt Institute for Surgery and Engineering, Vanderbilt University, USA

[7]Department of Neurological Surgery, Radiology, and Otolaryngology-Head and Neck Surgery, Vanderbilt University Medical Center, USA

[8]School of Mechanical Engineering, Weldon School of Biomedical Engineering, and Purdue Institute for Cancer Research, Purdue University, USA

[9]Livestrong Cancer Institutes and Departments of Biomedical Engineering, Diagnostic Medicine, and Oncology, The University of Texas at Austin, USA

[10]Department of Imaging Physics, The University of Texas MD Anderson Cancer Center, USA

**Running title.** Personalized forecasting of PCa during active surveillance.

**Keywords.** Prostate cancer, active surveillance, magnetic resonance imaging, computational oncology, biomechanistic modeling.

**Financial support**. G.L. reports a grant from the European Commission. J.S.H. reports a grant from NIH. M.I.M. reports a grant from NIH. H.G. reports a grant from NSF. T.E.Y. reports a grant from CPRIT. A.R. reports a grant from MIUR-PRIN.

**Corresponding author.** Guillermo Lorenzo, PhD. Department of Civil Engineering and Architecture, University of Pavia, Via Ferrata 3, 27100, Pavia, Italia. Email: guillermo.lorenzo@unipv.it, guillermo.lorenzo@utexas.edu

**Conflict of interest disclosure statement.** H.G., T.J.R.H., T.E.Y., A.R., and G.L. are listed as coinventors in a US patent application that has been filed by The University of Texas at Austin. The invention leverages part of the technology presented in this paper.

Word count (without methods): 4985

Word count (with methods): 8978

Figures and tables count: 9

# Abstract

Active surveillance (AS) is a suitable management option for newly-diagnosed prostate cancer (PCa), which usually presents low to intermediate clinical risk. Patients enrolled in AS have their tumor closely monitored *via* longitudinal multiparametric magnetic resonance imaging (mpMRI), serum prostate-specific antigen tests, and biopsies. Hence, the patient is prescribed treatment when these tests identify progression to higher-risk PCa. However, current AS protocols rely on detecting tumor progression through direct observation according to standardized monitoring strategies. This approach limits the design of patient-specific AS plans and may lead to the late detection and treatment of tumor progression. Here, we propose to address these issues by leveraging personalized computational predictions of PCa growth. Our forecasts are obtained with a spatiotemporal biomechanistic model informed by patient-specific longitudinal mpMRI data. Our results show that our predictive technology can represent and forecast the global tumor burden for individual patients, achieving concordance correlation coefficients ranging from 0.93 to 0.99 across our cohort ($n = 7$). Additionally, we identify a model-based biomarker of higher-risk PCa: the mean proliferation activity of the tumor ($p = 0.041$). Using logistic regression, we construct a PCa risk classifier based on this biomarker that achieves an area under the receiver operating characteristic curve of 0.83. We further show that coupling our tumor forecasts with this PCa risk classifier enables the early identification of PCa progression to higher-risk disease by more than one year. Thus, we posit that our predictive technology constitutes a promising clinical decision-making tool to design personalized AS plans for PCa patients.

# Significance statement

Personalization of a biomechanistic model of prostate cancer with multiparametric MRI data enables the prediction of tumor progression, thereby showing promise to guide clinical decision-making during active surveillance for each individual patient.

# Introduction

Prostate cancer (PCa) is the second most common type of cancer and the fifth leading cause of cancer death among men worldwide (1). The clinical management of PCa relies on two key strategies (2): regular screening and patient triage into risk groups. Regular screening of PCa is performed in men over fifty and consists of a digital rectal exam and the measurement of the serum level of the prostate specific antigen (PSA). IF the results these two tests are suggestive of PCa, then a multiparametric magnetic resonance imaging (mpMRI) scan will usually be performed to assess the prostate non-invasively and also guide a subsequent biopsy for histopathological confirmation (2, 3). Based on the data collected during this diagnostic stage, clinical protocols assign risk groups and associate specific management options to each of them (2). The definition of these risk groups usually relies on the serum PSA, the clinical stage (i.e., an estimate of tumor size and extent) according to the tumor-node-metastases (TNM) scale, and the biopsy Gleason score (GS), which is the gold standard histopathological marker of PCa aggressiveness and is defined upon the cancerous tissue architecture observed in biopsy samples or surgically-removed prostates (2, 4). Thanks to the current clinical management of PCa, most tumors are detected and successfully treated at early organ-confined stage, which usually poses a low to intermediate risk to the patient (5, 6). Although standard treatments for PCa (e.g., surgery, radiotherapy) exhibit high rates of overall and disease-free survival (2, 7, 8), many newly-diagnosed PCa cases are known to be indolent and may not produce any symptoms or require treatment for a long time. These patients are prone to potential treatment side-effects (e.g., incontinence, impotence) that can adversely impact their quality of life without improving longevity. Thus, overtreatment of PCa is a major concern in the current clinical management of the disease (3, 9, 10). Additionally, undertreatment of PCa due to limitations in diagnosis and management (e.g., biopsy sampling errors, mpMRI confounders, limited age and comorbidity assessment) constitutes another important clinical challenge, since it may result in rapid growth of aggressive tumors, treatment failure, and reduced survival (3, 9-11).

As an alternative to direct treatment after diagnosis, active surveillance (AS) is another standard-of-care management option that is suitable for many newly-diagnosed PCa cases exhibiting low to intermediate clinical risk (2, 3, 8, 12). In AS, patients are closely monitored *via* longitudinal mpMRI scans, digital rectal exams, PSA tests, and biopsies, such that treatment is delayed until these monitoring tests confirm PCa progression to higher risk. Thus, AS is widely regarded as an ideal clinical strategy to reduce the overtreatment of newly-diagnosed PCa (2, 3, 12). Additionally, the changes in longitudinal data collected during AS (e.g., mpMRI-observable features, PSA kinetics) can be leveraged to inform clinical decision-making and, hence, contribute to reduce treatment excesses and deficiencies (12, 13). However, current AS protocols largely rely on recommending monitoring tests either at fixed times (e.g., mpMRI scan potentially followed by biopsy every 6-36 months, PSA test every 3-12 months) or after the observation of clinical events suggesting tumor progression (e.g., imaging after a sustained PSA rise, biopsy after radiological worsening of an mpMRI-detected lesion; see Refs. (2, 12, 13)). This population-based, observational approach largely ignores the complex underlying tumor dynamics and the intrinsic heterogeneity of PCa between and within patients (3, 4, 13-15), complicates the design of personalized plans to ensure optimally controlled tumor monitoring, and hinders the early detection of PCa progression, which may occur between two consecutive and widely-spaced monitoring tests. Thus, the optimal timing and type of monitoring tests for each individual PCa case constitutes a central challenge in AS, along with the definition of triggering criteria to switch from monitoring to radical treatment (2, 3, 12, 13, 16). To address these unmet and timely demands in AS, we propose to leverage personalized computational predictions of PCa growth (17, 18).

Computational tumor forecasting is a technology that enables the estimation of the growth of tumors and their response to treatment on a personalized basis (18). These tumor forecasts are obtained by leveraging mathematical models based on differential equations or agent-based formulations that describe the main biological mechanisms underlying the development of these diseases and the effect of specific treatments (18, 19). By calibrating the model parameters governing tumor dynamics with patient-specific data, the ensuing personalized model enables the representation of observed tumor growth as well as the prediction of the future development of the tumor, which can be leveraged to guide monitoring strategies or adjust treatment regimens for each individual patient (18-26). Indeed, the increasing success of computational tumor forecasting has motivated its use to design digital twin technologies to systematically guide clinical decision making using a predictive, patient-specific approach (27). In particular, medical imaging data, such as mpMRI scans, provide spatially-resolved anatomical and physiological information about tumors that enable the personalization of spatiotemporal biomechanistic models based on partial differential equations, which can then render three-dimensional tumor forecasts within the patient's affected organ (18). While this type of models has been extensively used to predict the development and therapeutic response of other tumors (20-23, 28-31), there is a dearth of applications for PCa (17, 32-35). Instead, most biomechanistic models developed for PCa consist of time-resolved differential formulations primarily based on serum PSA dynamics that have been successfully used to predict response to specific treatments, especially in the advanced stage of the disease (24-26, 36). The abundance of these PSA models is primarily motivated by the extensive use of this biomarker to monitor PCa, while longitudinal mpMRI data are almost only available in AS. However, time-resolved models lack spatial information on PCa growth that can be key during AS, for example, to plan biopsies or to assess malignancy in prostate areas that may associate with worse prognosis (e.g., prostate capsule, regions adjacent to neurovascular bundles and seminal vesicles; see Refs. (2-4, 13)).

Here, we present a study of personalized forecasting of untreated PCa growth during AS using a spatiotemporal biomechanistic model informed by longitudinal mpMRI data collected for each individual patient. Toward this end, we develop a clinical-computational pipeline to segment the prostate and tumor, perform intra- and interscan registration of imaging data, build a three-dimensional virtual representation of the prostate geometry holding the longitudinal spatial changes of the tumor, calibrate a biomechanistic model of PCa growth, and perform patient-specific tumor forecasting (see Figure 1). First, we analyze whether our computational framework can represent PCa growth by informing the model with three imaging datasets for each patient. Then, we investigate whether we can predict PCa growth after informing the model with two imaging datasets by comparing a prediction of the personalized model against a third mpMRI dataset. Additionally, since PCa risk and AS eligibility heavily rely on GS (2, 12), we further analyze our personalized computer simulations to search for potential model-based predictive biomarkers of PCa progression to higher-risk disease.

## Materials and Methods

*Patient data*

Anonymized patient data were retrospectively collected at The University of Texas Health Science Center San Antonio (San Antonio, TX, USA) following an institutional review board-approved and HIPAA-compliant protocol, which did not require informed consent. The inclusion criteria were: (i) newly-diagnosed, untreated, organ-confined PCa managed with AS after diagnosis; (ii) availability of three mpMRI scans during AS; (iii) a visible tumor across all mpMRI scans; and (iv) histopathologically-confirmed PCa *via* biopsy and/or analysis of the surgically-removed prostate after AS. The cohort leveraged in this study consisted of seven patients ($n = 7$), each of them having one organ-confined tumor matching the inclusion criteria. The longitudinal mpMRI scans of each patient were collected over a period of median (range) of 3.4 (1.4, 4.9) years with interscan frequency of 1.4 (0.5, 2.6) years. For each patient, we define a relative time scale beginning at the time of the first mpMRI scan ($t = 0$). Although most biopsies were performed after an mpMRI scan, some patients also had biopsies before the first imaging session. Baseline GS at $t = 0$ was assigned according to the GS of the first biopsy performed within 3 months after the first mpMRI date. If there was no biopsy within this period but the patient had a posterior biopsy (i.e., > 3 months after the first mpMRI scan) and a pre-imaging biopsy, the first post-imaging GS was used as the baseline measurement only if it was identical to the pre-imaging GS value. If the patient had no biopsies after the first mpMRI scan, the most recent GS measured before the first mpMRI was used as baseline. All other post-imaging GS measurements were used for temporally-informed clinical risk stratification during AS. This process enabled the definition of GS at $t = 0$ for all patients and resulted in a total of 16 eligible GS measurements with a median (range) of 3 (1, 4) values per patient. These GS measurements were evenly distributed in two subgroups exhibiting a GS = 3+3 ($n = 8$) and a GS ≥ 3+4 ($n = 8$), which we denote as lower-risk and higher-risk PCa. This division enabled a balanced subgroup definition and is consistent with the fact that PCa with GS 3+3 is extensively recognized as eligible for AS, while there is debate regarding the inclusion in AS protocols of PCa cases exhibiting Gleason grade 4 (2, 12).

*MRI protocols*

The mpMRI datasets were collected using 3T scanners for all subjects. A TIM Trio scanner (Siemens Healthineers USA, Malvern, PA) housed at the Research Imaging Institute of The University of Texas Health Science Center San Antonio was used to obtain a baseline scan with a cardiac coil used on the pelvis (37, 38). Follow-up scans were performed following the standard-of-care using either a GE 750W (GE Healthcare USA, Chicago, IL) at the South Texas Veterans Healthcare System or a Siemens Skyra scanner (Siemens Healthineers USA, Issaquah, WA) at The University of Texas Health Science Center San Antonio. The mpMRI acquisition protocols included $T_1$-weighted, $T_2$-weighted, diffusion-weighted, and dynamic contrast-enhanced sequences (T1W, T2W, DW, and DCE, respectively). The main acquisition parameters were as follows: echo time <90 seconds, repetition time >3000 ms, slice thickness <4.0 mm without gap, field of view of 160-220 mm, and b-values of approximately 50, 400, 800, and 1400 s/mm$^2$ on all scanners. Apparent diffusion coefficient (ADC) maps were calculated from DW-MRI data using the standard monoexponential model (39).

*PCa histopathology*

All biopsies were performed using a transrectal ultrasound approach. After an mpMRI was performed, the regions of interest and prostate boundaries were marked in DynCAD software as part of the UroNav (Phillips Healthcare, USA, Cambridge, MA) biopsy system. During biopsy, once the prostate was scanned and registered with the MRI, three targeted biopsies were obtained from each region of interest and a standard 12-core prostate biopsy was also completed. Biopsy specimens underwent standard pathologic processing and were graded using the International Society of Urologic Pathology (ISUP) criteria and standard Gleason grade scoring. The same histopathological assessment process was used for standard biopsies before the first mpMRI scan and for surgically-removed prostate specimens.

*Preprocessing of imaging data*

The integration of the mpMRI data from each patient within our biomechanistic model of PCa growth required five preprocessing operations: (i) segmentation of the prostate and its zonal anatomy in each mpMRI dataset, (ii) intrascan registration of the T2W map to the ADC map for each mpMRI scan, (iii) segmentation of the tumor on each ADC map, (iv) interscan registration of the longitudinal mpMRI data and the segmentations for each patient, and (v) conversion of intratumoral ADC values to tumor cell density estimates. This computational pipeline was designed following previous efforts developed to support mpMRI-informed personalized forecasting for brain (29) and breast cancers (21, 28). For each mpMRI dataset, the segmentation of the prostate and the tumor, as well as the intrascan registration, were carried out using a combination of manual and semi-automatic tools available in 3DSlicer (40). We next briefly describe the preprocessing operations in the order they were implemented.

*Prostate segmentation.* We first resampled the T2W and ADC maps to isotropic resolution using linear interpolation. Then, the prostate and its central gland were manually segmented in the resampled T2W images and ADC maps using anatomical landmarks as reference (3, 4). The local anatomy of the prostate was completed by defining the peripheral zone as the region of the prostate outside the central gland.

*Intrascan registration.* To align the ADC map to the T2W map of each mpMRI dataset, we performed a rigid intrascan registration by leveraging the *General Registration* module of 3DSlicer and using the corresponding prostate segmentations for masking. Following this intrascan registration, we smoothed the T2W-based segmentations of the prostate and the central gland using the *Segmentation smoothing* module in 3DSlicer.

*Tumor segmentation.* The segmentation of the tumor consisted of a multistep, hierarchical procedure. First, we manually drew a gross segmentation of the tumor over the registered ADC map based on the tumor location provided in the mpMRI reports, which was confirmed by an experienced urologist who specializes in PCa (M.A.L.). Second, we manually delineated a volume of healthy tissue in the same region of the prostate where the tumor is primarily located (i.e., peripheral zone or central gland) with approximately the same volume as the gross segmentation of the tumor. We then calculated the mean ADC over this volume as an estimate of ADC in healthy tissue ($ADC_h$) using the *Segment Statistics* module in 3DSlicer. Third, we used the *threshold* tool in 3DSlicer to delineate the tumor core within the gross segmentation of the tumor, which was defined as the isovolume of ADC values with less than 70% of $ADC_h$. This value was selected after an analysis of five studies investigating the correlation of ADC and GS (41-45). These studies show that PCa exhibits lower ADC than healthy prostatic tissue and that ADC progressively decreases as the GS increases. This trend seems to plateau for higher GS values, which aligns with a higher tumor cell density approaching tissue carrying capacity in those tumors (46-48). By fitting a hyperbolic tangent function to the average of the ADC ratios ($ADC/ADC_h$) obtained from the mean $ADC_h$ and mean GS-specific ADC values in the five studies (41-45), we determined that the 70% of $ADC_h$ was a sufficient threshold to identify mpMRI-visible tumors (see Figure 2A). This fitting was performed using the Curve Fitting Toolbox in MATLAB (R2021b; The Mathworks, Natick, MA, USA) and defining a continuous GS scale ranging from 0 to 10, where values between 0 and 2 represent healthy tissue and pretumoral lesions (4). Additionally, the choice of the hyperbolic tangent includes a slow decrease of ADC in pretumoral lesions and low GS tumors, which makes them practically indistinguishable from healthy tissue and aligns with the fact that mpMRI-observable PCa generally exhibits GS ≥ 3+3 at diagnosis (3, 14, 16). Supplementary Methods S1 provides further details about this hyperbolic tangent fit. Finally, the resulting tumor core was expanded to define the final tumor segmentation by leveraging the *margin* tool in 3DSlicer. This technique was applied within the gross tumor segmentation and using a margin of 2 to 4 mm (equivalent to a kernel from 3×3×3 to 5×5×5 voxels, depending on the tumor size), such that the tumor border lied over larger ADC values (e.g., $\gtrsim$ 90% of $ADC_h$). Hence, this margin facilitated a smooth transition of ADC values from the tumor core to the neighboring healthy tissue that matches the natural solution of our biomechanistic model.

*Interscan registration.* For each patient, the three mpMRI datasets collected during AS and their corresponding segmentations were co-registered to a common data frame, which was set as that of the first mpMRI scan (see Figure 1C). This interscan registration was carried out using a biomechanically-constrained, deformable image registration algorithm to control for bulk deformations of the prostate anatomy between scan dates while preserving imaging features associated with longitudinal disease-related changes. The registration process consisted of a rigid registration of the T2W-based prostate segmentations followed by a deformable linearized iterative boundary reconstruction algorithm. A robust rigid alignment of prostate anatomy was established by a semiautomatic, salient feature weighted, iterative closest point algorithm (49). Salient features were marked from base to apex along the posterior, left, and right aspects of the prostate to aid initial anatomical alignment. The linearized iterative boundary reconstruction algorithm (50) was then employed to solve for an optimal distribution of external forces applied to the outer boundary of the prostate to maximize alignment between source and target prostate geometries while preserving biomechanical consistency. Briefly, a series of 90 control points were uniformly distributed over the boundary of the prostate *via* k-means clustering, which partitioned the prostate surface into an equal number of Voronoi control surfaces. Linear elastic biomechanical perturbations of control points to 5-mm displacements in each Cartesian direction were simulated using a finite-element model of the prostate. Near-field displacement responses to these control point perturbations were relaxed over their respective control surfaces according to the Saint-Venant principle by re-equilibrating the local boundary forces within the active control surface against far-field displacements resulting from the point perturbation. These relaxed perturbation responses establish a superposed basis of elastic deformation responses to locally decomposed mechanical loads that are regionally applied over the prostate boundary. A linear combination of this superposed basis subsequently encodes a parameterized deformation response to a continuous, spatially varying distribution of mechanical forces applied over the prostate surface. This parameterized model for whole-organ prostate deformations was then optimized to minimize the distance between the source and the target prostate segmentation boundaries with regularization by a strain energy penalty function to resolve discrepant external forces applied to the prostate across mpMRI scans. Further details are found in (50). This registration method filters out purely biomechanical deformation effects between imaging dates while preserving the effect of non-elastic physiological changes associated with tumor growth that may occur during AS. As a final step, the interscan registration was also applied to the T2W data, the ADC maps, and the tumor segmentations of each patient. After the interscan registration, we updated the reference prostate segmentation (i.e., that of the first mpMRI scan) to prepare it for integration within our modeling framework by subtracting the urethral region. To do this, we used the *eraser* tool in 3DSlicer with a spherical geometry and 4-mm diameter. Anatomical landmarks in the T2W data were used for reference.

*Conversion of intratumoral ADC values to tumor cell density estimates.* Since ADC has been shown to decrease as tumor cell density increases in higher GS tumors (41-48), the ADC values within each tumor segmentation were converted to normalized tumor cell density by leveraging a formulation that has been successfully employed to represent the inverse relationship between ADC and tumor cell density in other solid tumors (18, 21, 28, 29), as follows:

$$\widehat{N}(\boldsymbol{x},t) = \frac{N(\boldsymbol{x},t)}{\theta} = \frac{ADC_h - ADC(\boldsymbol{x},t)}{ADC_h - ADC_{min}}. \tag{1}$$

In Eq. (1), $\widehat{N}(\boldsymbol{x},t)$ is the normalized tumor cell density at position $\boldsymbol{x}$ and time $t$ (i.e., $0 \leq \widehat{N}(\boldsymbol{x},t) \leq 1$), which is defined as the ratio of tumor cell density $N(\boldsymbol{x},t)$ to the tissue carrying capacity $\theta$ (i.e., the maximally admissible tumor cell density in the prostate, $0 \leq N(\boldsymbol{x},t) \leq \theta$). Additionally, $ADC_h$ is the ADC in the healthy tissue as defined above, $ADC(\boldsymbol{x},t)$ is the ADC at position $\boldsymbol{x}$ and time $t$ within the tumor, and $ADC_{min}$ is the minimum ADC observable within a tumor. For each ADC map, the latter was estimated from the lower horizontal asymptote of the hyperbolic tangent function in Figure 2A as $ADC_{min}/ADC_h = 0.25$. Figure 2B further provides the mapping through Eq. (1) of the hyperbolic tangent function describing the changes in ADC ratio with respect to GS, thereby yielding the changes in normalized tumor cell density with respect to GS within our modeling framework. Since $ADC_h$ and $ADC_{min}$ are fixed estimates, ADC values above $ADC_h$ and below $ADC_{min}$ were truncated to render a normalized tumor cell density of 0 and 1, respectively. This operation avoided unphysical negative tumor cell densities and values over the carrying capacity. Thus, according to Eq. (1) and the plots in Figure 2, the segmentation of the tumor core using values below 70% $ADC_h$ corresponds to normalized tumor cell density values $\widehat{N} \geq 0.4$, and, after the margin expansion, the border of the resulting tumor segmentation has values $\widehat{N} \lesssim 0.15$.

*Biomechanistic model*

We model PCa growth during AS in terms of the spatiotemporal dynamics of the tumor cell density, $N = N(\boldsymbol{x},t)$, using a reaction-diffusion partial differential equation, as follows:

$$\frac{\partial N}{\partial t} = \nabla \cdot (D \nabla N) + \rho N \left(1 - \frac{N}{\theta}\right). \tag{2}$$

Eq. (2) is commonly known as the Fisher-Kolmogorov equation (51). Many patient-specific tumor forecasting studies have directly relied on this equation or leveraged it as a basis to build more sophisticated models (18, 21-23, 28-31). The right-hand side of Eq. (2) describes the dynamics of the tumor cell density as a combination of two mechanisms: tumor cell mobility, which is represented by a diffusion process governed by the tumor cell diffusivity ($D$), and tumor cell net proliferation, which is modeled with a logistic term controlled by the net proliferation rate ($\rho$) and the tissue carrying capacity ($\theta$). The tumor cell net proliferation rate encompasses the balance of tumor cell proliferation and death. In this work, we assume that the carrying capacity is constant for all patients, which enables posing our model in terms of the normalized tumor cell density ($\widehat{N} = N/\theta$) by dividing both sides of Eq. (2) by $\theta$:

$$\frac{\partial \widehat{N}}{\partial t} = \nabla \cdot \left(D \nabla \widehat{N}\right) + \rho \widehat{N}\left(1 - \widehat{N}\right). \tag{3}$$

Hence, once we calibrate the values of parameters $D$ and $\rho$ in Eq. (3) that best explain PCa dynamics as observed in the longitudinal mpMRI measurements of normalized tumor cell density collected during AS for each patient, we can obtain personalized forecasts of tumor growth. Figure 3 illustrates how tumor cell mobility and net proliferation contribute to PCa dynamics according to the models in Eqs. (2) and (3). As shown in Figure 3, these models are posed within the patient's prostate geometry, which corresponds to the T2W-based segmentation from the first mpMRI scan. Furthermore, the model is defined over a temporal interval $[0, T]$, where $T$ is a specific time horizon (e.g., for model calibration or tumor forecasting for each patient). To complete the model, we need to define boundary conditions over the prostate surface. Since the PCa cases eligible for this study were organ-confined, we set zero-flux boundary conditions (i.e., $\nabla N \cdot \boldsymbol{n} = 0$ or, equivalently, $\nabla \widehat{N} \cdot \boldsymbol{n} = 0$; where $\boldsymbol{n}$ is an outward unit vector orthogonal to the prostate boundary). Additionally, the initial conditions for Eq. (3), $\widehat{N}(\boldsymbol{x}, 0)$, were always defined with the normalized tumor cell density map obtained from the first mpMRI scan.

To calibrate the model and obtain PCa predictions for each patient, we performed computer simulations of our model by solving Eq. (3) numerically. These computer simulations constitute an *in silico* representation of personalized PCa growth over the patient's prostate anatomy and the temporal interval $[0, T]$, which we can compare to mpMRI data at scan times. We leveraged isogeometric analysis (IGA), which is a high-fidelity generalization of the finite element method that uses highly smooth functions to discretize the geometry and the solution of partial differential equations (18, 52). In particular, we discretized Eq. (3) in space using a standard isogeometric Bubnov-Galerkin method that relied on a three-dimensional $C^1$ quadratic non-uniform rational B-spline (NURBS) functional space (18, 33, 52). This method requires the construction of a patient-specific mesh of the prostate using the same function space. Toward this end, we used a parametric mapping method (17, 32), which deformed a reference torus mesh to match the final prostate segmentation from the first mpMRI scan of each patient. The original torus and prostate meshes had a polar discretization with 32 elements in the circumferential directions and 8 elements in the radial direction. Each prostate mesh was subsequently refined using standard knot insertion (52) for numerical accuracy in the computer simulations of our biomechanistic model. Hence, the final patient-specific prostate meshes had 128 elements in the circumferential directions and 32 elements in the radial direction. The tumor segmentations and corresponding normalized tumor cell density maps were then $L^2$-projected over the prostate mesh for each patient (17, 32, 33). Furthermore, the time interval $[0, T]$ was discretized with a constant time step $\Delta t = 1$ day for all patients, which sufficed to capture the spatiotemporal dynamics observed in the patients of this study. We integrated in time using the generalized-$\alpha$ method (33, 52). This iterative algorithm produces a series of nonlinear systems of equations in each time step, which we linearized with the Newton-Raphson method. The ensuing linear systems were solved utilizing the generalized minimal residual method (GMRES). Further details on the implementation of these numerical methods to solve our biomechanistic model are available in (17, 18, 32, 33, 52).

*Model calibration*

The personalized parameterization of our biomechanistic model consisted of finding the values of the parameter set $(D, \rho)$ that minimize the relative squared difference between the model predictions and the corresponding imaging-based measurements of normalized tumor cell density at the times of the mpMRI scans used to inform the model. This calibration problem can be formulated as:

$$(D, \rho) = \arg\min \left( \sum_{i=1}^{n_s} \frac{\int_\Omega \left(\widehat{N}_d(\boldsymbol{x}, t_i) - \widehat{N}_m(\boldsymbol{x}, t_i, D, \rho)\right)^2 d\Omega}{\int_\Omega \left(\widehat{N}_d(\boldsymbol{x}, t_i)\right)^2 d\Omega} \right), \quad (4)$$

where $n_s$ is the number of mpMRI scans used for model calibration, $\Omega$ represents the three-dimensional geometry of the patient's prostate, and $\widehat{N}_d(\boldsymbol{x}, t_i)$ and $\widehat{N}_m(\boldsymbol{x}, t_i, D, \rho)$ denote the normalized tumor cell density measured from imaging data and calculated from the model with parameters $D$ and $\rho$ at the scan times $t_i$ ($i = 1, \ldots, n_s$), respectively. Hence, Eq. (4) defines a nonlinear least-squares problem to find the patient-specific parameter values $(D, \rho)$, which we solved by leveraging the Gauss-Newton method (53). The initial guess for the parameters was set as $D = 5 \cdot 10^{-3}$ mm²/day and $\rho = 2 \cdot 10^{-3}$ 1/day, while the admissible values of the parameters were constrained to $D \in [1.0 \cdot 10^{-6}, 10]$ mm²/day and $\rho \in [1.0 \cdot 10^{-6}, 1]$ 1/day.

Model performance during calibration and forecasting was assessed with a panel of global and local metrics. The global metrics were the tumor volume ($V_T$) and the total tumor cell volume ($V_N$), which are defined as:

$$V_T = \int_{\Omega_T} d\Omega \quad (5)$$

$$V_N = \int_{\Omega_T} \widehat{N} d\Omega \quad (6)$$

In Eqs. (5) and (6), $\Omega_T$ is the three-dimensional tumor region where $\widehat{N} \geq \widehat{N}_{th}$ with $\widehat{N}_{th}$ being a threshold value to systematically identify tumor tissue both in the imaging data and the model forecasts. Hence, this threshold facilitated the automatic delineation of a tumor region in the imaging data, for which the values over the tumor segmentation border were $\widehat{N} \lesssim 0.15$, and from the model simulations, which provided a continuous $\widehat{N}$ map within the range [0,1]. For this purpose, we set $\widehat{N}_{th} = 0.15$ in all the calculations presented in this work. The global metrics $V_T$ and $V_N$ were calculated patient-wise in each calibration scenario, and model-data agreement was assessed *via* the Pearson and concordance correlation coefficients over the cohort (PCC and CCC, respectively). The local metrics employed to analyze the performance of the model were: the Dice similarity coefficient (DSC) as well as the root mean squared error (RMSE), the local PCC, and the local CCC between the tumor cell density maps calculated from the imaging data and leveraging our biomechanistic model. These local metrics were calculated patient-wise and we analyzed their distribution using boxplots.

*Model-based biomarkers of high-risk PCa*

We analyzed the personalized computer simulations of our model to find model-based biomarkers of high-risk PCa. Toward this end, we first calculated the values of a panel of six candidate markers at the times of histopathological assessment of the tumors in our patient cohort, and then compared their values between the subgroups of low-risk and high-risk PCa (i.e., GS = 3+3 and GS ≥ 3+4, respectively). These candidate markers were calculated using the model simulations obtained with the personalized parameters from the global calibration study, since this scenario informed the model with the maximum amount of data available for each patient in this study. The list of candidate markers was designed based on previous tumor forecasting studies (22, 23, 28-32, 54) and our analysis of the personalized predictions in this work. The six candidate markers were: the prostate volume ($V_P$), the tumor volume ($V_T$), the total tumor cell volume ($V_N$), the mean normalized tumor cell density ($\overline{N}$), the total tumor index ($N_T$), and the mean proliferation activity of the tumor ($A_p$). The prostate volume and the last three markers were calculated as follows:

$$V_P = \int_\Omega d\Omega \qquad (7)$$

$$\overline{N} = \frac{\int_{\Omega_T} \widehat{N} d\Omega}{\int_{\Omega_T} d\Omega} = \frac{V_N}{V_T} \qquad (8)$$

$$N_T = \overline{N}\frac{V_T}{V_P} \qquad (9)$$

$$A_p = \int_{\Omega_T} \rho \widehat{N}(\boldsymbol{x},t)\left(1 - \widehat{N}(\boldsymbol{x},t)\right) d\Omega \qquad (10)$$

The prostate volume and the total tumor index defined in Eqs. (7) and (9), respectively, were included to account for the organ size effect on tumor dynamics, since high-volume prostates enlarged by concomitant benign prostatic hyperplasia (BPH) have been suggested to exhibit a lower risk of high-grade PCa (32, 55). Additionally, the mean proliferation activity of the tumor ($A_p$) was defined based on the work presented in (54).

Any candidate marker exhibiting a statistically significant difference between the low-risk and high-risk PCa subgroups was considered an eligible model-based biomarker of high-risk PCa. We analyzed the performance of these biomarkers by building univariate logistic regression classifiers and using receiver operating characteristic (ROC) curve analysis. We further built bivariate logistic regression classifiers including a maximum of two markers from the candidate list, such that at least one was identified as a biomarker. We did not consider logistic regression classifiers with more than two input markers to avoid overfitting given the reduced size of the histopathological measurements in the cohort ($n$ = 16). All logistic regression classifiers were built using the function *glmfit* from the Statistics and Machine Learning Toolbox in MATLAB. The area under the ROC curve was calculated numerically with the trapezoidal rule and the optimal performance point was determined using the minimum distance to the upper left corner (i.e., where sensitivity and specificity are maximal).

Given the small size of the cohort in this work ($n = 7$), we did not partition it to define a patient subgroup to train the classifiers and another group to validate them. Nevertheless, to investigate their predictive performance, we analyzed them using the personalized computer simulations from the fitting-forecasting study. In particular, we analyzed whether the best performing classifier, which was trained with markers calculated with personalized simulations from the global calibration study, was able to anticipate the development of high-risk disease for each patient when the model was only informed by the first two mpMRI scans.

*Statistical methods*

We leveraged Wilcoxon rank-sum and signed-rank tests to compare the local metrics of model performance calculated at the times of the second and third mpMRI scans in both the global calibration scenario and the fitting-forecasting study. These two statistical tests were also used to investigate differences in the model parameters as well as global and local metrics of model performance between the global calibration scenario and the fitting-forecasting study. Additionally, the Wilcoxon rank-sum test was also used to identify differences between the candidate markers of high-risk PCa between the low-risk and high-risk PCa subgroups. The two types of statistical tests were performed using the functions *ranksum* and *signrank* from the Statistics and Machine Learning Toolbox in MATLAB. In the Results section, we specify when we use each type of test and whether it is two-tailed or one-tailed for each statistical analysis. The level of significance for all statistical tests was set to 5%.

*Data availability*

The raw data leveraged in this study was generated at The University of Texas Health Science Center San Antonio (San Antonio, TX, USA). Raw data are not publicly available under the IRB data usage agreement for this study, but the derived data that support the findings presented in this work are available from the corresponding author upon reasonable request.

# Results

*The biomechanistic model represents PCa growth during AS for each individual patient*

We first performed the global calibration study to assess the ability of our computational framework to represent untreated PCa growth during AS. In this study, the biomechanistic model was informed by the three mpMRI scans available for each patient ($n = 7$). The reference prostate geometry from the first mpMRI scan had median (range) volume of 36.4 (18.5, 67.3) cc. The imaging-based segmentation of the tumors over the first, second, and third mpMRI datasets resulted in a tumor volume ($V_T$) of 0.12 (4.4·10$^{-3}$, 1.26) cc, 0.28 (0.06, 4.44) cc, and 0.49 (0.14, 7.12) cc, respectively. The corresponding total tumor cell volumes ($V_N$) were 0.07 (1.5·10$^{-3}$, 0.91) cc, 0.13 (0.02, 2.45) cc, and 0.31 (0.06, 3.89) cc. The global calibration study resulted in a distribution of tumor cell diffusivity coefficients ($D$) and net tumor cell proliferation rates ($\rho$) of 1.26·10$^{-3}$ (5.74·10$^{-4}$, 5.35·10$^{-3}$) mm$^2$/day and 1.86·10$^{-3}$ (1.09·10$^{-3}$, 5.13·10$^{-3}$) 1/day, respectively. After initializing our biomechanistic model with the normalized tumor cell density maps from the first mpMRI scan and corresponding calibrated parameters for each patient, the ensuing personalized simulations resulted in tumor volumes ($V_T$) of 0.33 (0.04, 5.44) cc and 0.59 (0.15, 7.69) cc at the time of the second and third mpMRI scans, respectively. The corresponding model-calculated total tumor cell volumes ($V_N$) were 0.13 (0.01, 2.28) cc and 0.23 (0.04, 3.37) cc.

Figure 4 illustrates the three-dimensional imaging measurements and calculations of the tumor geometry and the normalized tumor cell density map obtained with the personalized model for four patients at the second and third imaging timepoints. The corresponding results for the other three patients are provided in Supplementary Figure S1. Additionally, Figures 5A and 5B provide unity plots comparing the imaging measurements and the model calculations of the tumor volume ($V_T$) and the total tumor cell volume ($V_N$). Considering the second mpMRI scan for each patient, the PCC and CCC for $V_T$ were 0.99 and 0.97, respectively. The corresponding PCC and CCC for $V_N$ were 0.99 and 0.98. At the time of the third mpMRI scan, the PCCs and CCCs of both $V_T$ and $V_N$ were ≥ 0.99 in all cases. Figures 5C-5F further provide the distribution of the local metrics assessing the pointwise agreement between the imaging-based and the model-calculated normalized tumor cell density maps ($\widehat{N}(\boldsymbol{x},t)$) across the patient cohort ($n = 7$). At the second imaging timepoint, we obtained a DSC of 0.81 (0.66, 0.85), an RMSE of 0.022 (0.008, 0.071), a local PCC of 0.50 (0.19, 0.68), and a local CCC of 0.49 (0.18, 0.67). The corresponding distributions of DSC, RMSE, local PCC, and local CCC at the time of the third mpMRI scan were 0.77 (0.64, 0.85), 0.044 (0.016, 0.107), 0.28 (0.18, 0.63), and 0.28 (0.17, 0.61). No significant differences were detected between the values of each local metric at the second and third imaging timepoints under Wilcoxon rank-sum testing ($p > 0.05$). Nevertheless, two-sided Wilcoxon signed-rank tests at the time of the third mpMRI scan detected significantly higher RMSE ($p = 0.016$) and significantly lower DSC, local PCC, and local CCC ($p = 0.047$ in all three cases).

*Personalized computational forecasts predict untreated PCa growth*

To investigate the predictive performance of our computational framework, we carried out a fitting-forecasting study: for each patient, we first initialized our biomechanistic model with the data from the first mpMRI, we then calibrated the parameters using the data from the second mpMRI dataset, and we finally validated a personalized forecast of PCa growth against the imaging data collected at the third imaging timepoint. The median (range) of the tumor cell diffusivity coefficients ($D$) and net tumor cell proliferation rates ($\rho$) obtained in the fitting-forecasting study for each patient ($n = 7$) were $1.40 \cdot 10^{-3}$ ($5.35 \cdot 10^{-4}$, $4.18 \cdot 10^{-3}$) mm²/day and $2.41 \cdot 10^{-3}$ ($7.84 \cdot 10^{-4}$, $5.15 \cdot 10^{-3}$) 1/day, respectively. The ensuing patient-specific model simulations resulted in tumor volumes ($V_T$) of 0.29 (0.04, 4.98) cc and 0.45 (0.16, 6.95) cc at the second and third imaging timepoints, respectively. The corresponding total tumor cell volumes ($V_N$) were 0.12 (0.01, 2.20) cc and 0.20 (0.05, 3.52) cc.

Figure 6 presents the model-predicted tumor geometry and normalized tumor cell density map obtained in the fitting-forecasting study at the times of the second and third mpMRI scans for the same four patients shown in Figure 4 in the global calibration study, along with the corresponding imaging-based measurements. Similar results for the other three patients are provided in Supplementary Figure S2. Additionally, the unity plots in Figures 7A and 7B compare the model predictions of the tumor volume ($V_T$) and the total tumor cell volume ($V_N$) to their corresponding imaging measurements. At the time of the second mpMRI scan (i.e., the calibration horizon), the PCC and CCC values for the model-data agreement of $V_T$ and $V_N$ were ≥ 0.99 in all cases. The PCC and the CCC for $V_T$ resulted in 0.97 and 0.96 at the third imaging timepoint (i.e., the forecasting horizon), respectively. The corresponding values of the PCC and the CCC for $V_N$ at the third mpMRI date were 0.93 in both cases. Figures 7C-7E also present the distributions of the local metrics quantifying the pointwise agreement between the imaging measurements and model forecasts of the normalized tumor cell density maps ($\widehat{N}(x,t)$) over the patient cohort ($n$ = 7). At the second imaging timepoint, we obtained a DSC of 0.82 (0.67, 0.85), an RMSE of 0.022 (0.008, 0.068), a local PCC of 0.55 (0.19, 0.67), and a local CCC of 0.55 (0.18, 0.66). The corresponding values of the DSC, RMSE, local PCC, and local CCC at the forecasting horizon were 0.76 (0.65, 0.85), 0.048 (0.016, 0.137), 0.37 (0.26, 0.61), and 0.33 (0.24, 0.60), respectively. Two-sided Wilcoxon rank-sum testing resulted in no significant differences ($p$ > 0.05) between the values of the local metrics at the times of the second and third mpMRI scans, although the DSC values were identified as significantly higher at the calibration horizon with respect to the forecasting timepoint under a one-sided Wilcoxon rank-sum test ($p$ = 0.049). According to two-sided Wilcoxon signed-rank tests, model forecasts at the third imaging timepoint exhibited a significantly higher RMSE ($p$ = 0.016) and significantly lower DSC, local PCC, and local CCC ($p$ = 0.016, 0.047, and 0.047, respectively).

No significant differences were detected between the values of the model parameters ($D, \rho$) obtained in the global calibration and the fitting-forecasting study under both Wilcoxon rank-sum and signed-rank testing ($p$ > 0.05). Nevertheless, we observed a tendency toward lower tumor cell diffusivity coefficients ($D$) and higher net tumor cell proliferation rates ($\rho$) in the fitting-forecasting study for each patient under one-sided Wilcoxon signed-rank tests ($p$ = 0.055). Additionally, no Wilcoxon rank-sum test identified significant differences between the local model-data agreement metrics from the global calibration and the fitting-forecasting studies at either the second or third imaging timepoints ($p$ > 0.05). However, two-sided Wilcoxon signed-rank testing resulted in significantly lower RMSE values at the third mpMRI date in the global calibration study ($p$ = 0.016) and at the second imaging timepoint in the fitting-forecasting study ($p$ = 0.016). Furthermore, the DSC at the second scan timepoint was significantly higher in the fitting-forecasting study under a one-sided Wilcoxon signed-rank test ($p$ = 0.039).

*Model-based biomarkers enable early identification of progression to high-risk PCa*

To search for potential model-based biomarkers of high-risk PCa, we investigated six quantities of interest that were calculated at the times of histopathological assessment of each patient's tumor ($n$ = 16) using the personalized model simulations from the global calibration study. The median (range) of the prostate volume ($V_P$), the tumor volume ($V_T$), the total tumor cell volume ($V_N$), the mean normalized tumor cell density ($\overline{N}$), the total tumor index ($N_T$), and the mean proliferation activity of the tumor ($A_p$) in the low-risk PCa group (i.e., GS = 3+3; $n$ = 8) were 37.7 (21.8, 39.5) cc, 0.17 (0.01, 5.09) cc, 0.07 (1.5·10$^{-3}$, 2.12) cc, 0.39 (0.26, 0.52), 2.31·10$^{-3}$ (3.82·10$^{-5}$, 5.36·10$^{-2}$), 3.61·10$^{-4}$ (2.28·10$^{-4}$, 4.94·10$^{-4}$) 1/day. The corresponding median (range) values in the high-risk PCa subgroup (i.e., GS ≥ 3+4, $n$ = 8) were 34.0 (18.5, 67.3) cc, 0.58 (0.12, 5.90) cc, 0.19 (0.05, 2.58) cc, 0.39 (0.30, 0.54), 9.61·10$^{-3}$ (1.69·10$^{-3}$, 6.53·10$^{-2}$), 3.88·10$^{-4}$ (3.57·10$^{-4}$, 1.07·10$^{-3}$) 1/day. Figures 8A-8F plot the distribution of these candidate markers in the low-risk and high-risk PCa subgroups. Only the mean proliferation activity of the tumor ($A_p$) was significantly larger in high-risk PCa according to a one-sided Wilcoxon rank-sum test ($p$ = 0.041). Figure 8G shows the ROC curve for a univariate logistic classifier of high-risk PCa that was constructed using the mean proliferation activity of the tumor ($A_p$). The AUC of this ROC curve is 0.77 and the optimal performance point operates at 75% sensitivity and specificity. Additionally, Figure 8G also shows the ROC curve for the best performing bivariate logistic classifier that could be built using the mean proliferation activity of the tumor ($A_p$) and one of the other markers, which resulted to be the total tumor index ($N_T$). The AUC for this ROC curve is 0.83 and the optimal performance point also operates at 75% sensitivity and specificity.

To investigate whether the bivariate logistic classifier would anticipate the detection of high-risk PCa at the time of the second mpMRI scan (i.e., once the biomechanistic model is fully calibrated for each patient, and we can perform a personalized forecast), we tested it in the fitting-forecasting scenario. The results of this analysis for the four patients in Figures 4 and 6 are shown in Figure 9, while the corresponding results for the other three patients are reported in Supplementary Figure S3. These figures further compare the performance of the classifier using the personalized model simulations from the global calibration scenario (i.e., those used for classifier training). The PCa cases in Figures 9A and 9D were consistently classified as low-risk and high-risk PCa, respectively, in both the global calibration and the fitting-forecasting study, thereby matching the initial histopathological assessment available for these tumors. Similar results were obtained for the patients with high-risk PCa in Figure 9B and Supplementary Figure S3A. For the latter, our modeling framework also anticipated the progression from GS 4+3 to GS 4+4 using the personalized predictions at the second imaging timepoint, which was 410 days (i.e., ~1.1 years) earlier than the final histopathological assessment at surgery. Importantly, although the early biopsies of the tumor in Figure 9C report a low-risk case, the personalized biomechanistic model predictions (calculated at the second imaging timepoint) consistently classified this tumor as high risk and the global calibration (calculated at the third imaging timepoint) further confirmed PCa progression right after the second scan. For this patient, our model anticipated the detection of PCa progression at the time of the second mpMRI, which was 1,011 days (i.e., ~2.8 years) before the final histopathological assessment after surgery. Similar results are obtained for the PCa case in Supplementary Figure S3B, for which tumor progression was detected 677 days (i.e., ~1.9 years) before surgery. Finally, the low-risk patient in Supplementary Figure S3C was consistently identified as high-risk with the predictions of the fitting-forecasting study, but assimilation of the third mpMRI dataset in the global calibration scenario correctly classified the tumor as low-risk.

## Discussion

The majority of newly-diagnosed PCa patients are estimated to exhibit low or intermediate-risk PCa at diagnosis (5, 6), for which AS is a standard clinical option (2, 3, 12). To optimize patient monitoring during AS and rationalize the treatment of PCa, it is of utmost importance to accurately identify the patients with indolent disease, who can continue to reap the benefits of AS and avoid treatment, as well as the patients with progressing PCa, who may require immediate treatment. Nevertheless, current AS protocols largely rely on assessing PCa status at standardized frequencies that are set upon the average dynamics of the disease observed in large clinical studies (2, 12, 13). This population-based, observational approach cannot anticipate PCa progression and does not enable an optimal surveillance of the unique growth dynamics of each patient's tumor. To address these limitations of AS, we propose to use personalized computational forecasts of PCa growth obtained with a spatiotemporal biomechanistic model informed by the longitudinal mpMRI data collected during AS for each individual patient. Hence, our forecasting technology ultimately aims at advancing AS from its current observational, population-based standard toward a predictive, patient-specific paradigm.

Our PCa forecasts primarily consist of the prediction of spatiotemporal tumor growth in terms of normalized tumor cell density maps over the three-dimensional anatomy of the patient's prostate. In this work, we have shown that our approach can represent and predict PCa growth according to global metrics of tumor volume and cellularity (see Figures 5 and 7). This performance is analogous to that reported in prior tumor forecasting studies (18, 28, 29). Our results also show that our computational framework exhibits a promising potential to capture the evolving three-dimensional morphology of the tumor within the patient's prostate (see Figures 4 and 6). This information can be central to the precise planning of radical treatments, such as surgery or radiotherapy (17, 18, 22, 23, 29). Additionally, computational predictions of the three-dimensional tumor morphology within the patient's prostate can be leveraged to guide biopsies and, therefore, reduce the risk of underestimation of GS with respect to the histopathology analysis of the surgically-removed prostate (2, 11). Furthermore, despite the simplicity of the biomechanistic model leveraged in this study, our results show a promising trend in the local agreement between imaging measurements and model calculations of the normalized tumor cell density maps (see Figures 4-7), which is comparable to other tumor forecasting studies (18, 23, 28, 29, 31).

Our patient-specific forecasts further enable the prediction of PCa risk by leveraging a set of biomarkers calculated from normalized tumor cell density maps obtained from the personalized model. In particular, we found that the mean proliferation activity of the tumor can be used as a promising biomarker of higher-risk PCa, whose classifying performance improved in combination with the total tumor index (see Eqs. (9) and (10)). Tumor cell proliferation has been found to be a central mechanism driving tumor growth and treatment response in previous biomechanistic modeling studies, and model-based biomarkers of proliferation activity have been found to correlate with tumor progression and unfavorable pathological outcomes (22, 25, 26, 30, 54, 56). In the case of PCa, previous clinical studies have also shown that a high proliferation activity has been associated with higher GS as well as with poorer prognosis, treatment outcomes, and survival (57, 58). Furthermore, the total tumor index combines three key variables in the analysis of PCa growth in the clinical literature: cellularity (measured *via* the average normalized tumor cell density), tumor volume, and prostate volume (see Eq. (9)). While the first two have been directly associated with higher-risk PCa (4, 12, 13, 41-48), the latter has been suggested to have an inverse correlation (i.e., such that larger prostates tend to harbor tumors with more pathologically favorable features; see Refs. (32, 55)). Thus, the model-based quantities selected in the present study to construct a predictive classifier of high-risk PCa agree with previous analyses of the driving mechanisms of tumor growth in modeling studies, as well as observations of PCa biology and clinical progression in the literature. These results along with the promising performance of the bivariate classifier to predict progression at the time of the second mpMRI scan suggest that our forecasts of PCa risk could contribute to guide fundamental clinical decisions in AS, such as the frequency of monitoring tests and the best timing to direct a patient with progressing PCa to definitive treatment. For example, should the PCa risk prediction reveal that a tumor will be at low risk at the one to three-year horizon, then the interscan time can be extended to monitor the tumor. Conversely, if the forecast of PCa risk reveals a trend toward progression, the subsequent scans could be performed earlier to confirm this pathological event and promptly treat the tumor.

Despite the promising results in personalized forecasting of PCa growth during AS observed in this work, we also acknowledge several limitations. First, we employed a small patient cohort ($n = 7$). Beyond statistical limitations (e.g., potential data correlation and low statistical power), this small cohort is not representative of the vast intratumoral and intertumoral heterogeneity of the newly-detected PCa cases that are eligible for AS. Additionally, the limited size of the cohort also prevented us from training and validating the PCa risk classifiers in different patient subgroups. Thus, future PCa forecasting studies should involve a larger cohort with more diversity of tumors, for example, in terms of location, number and size of lesions, prostate volumes, and GS changes during AS (e.g., stable indolent disease, early versus late progression, overall GS values; see Refs. (2, 11-15)). Second, we used both biopsy and surgically-based measurements of GS to train our high-risk PCa classifiers. However, PCa biopsies may underestimate GS mainly due to sampling bias and the incomplete understanding of imaging-histopathology data correlation (3, 11-13, 16, 48). Thus, a more robust training of our classifiers could be achieved by exclusively leveraging GS measurements on surgically removed prostate specimens paired to recent pre-surgery mpMRI scans. Third, our personalized forecasting approach does not account for uncertainty quantification and requires at least two mpMRI datasets to initialize and calibrate the model. We employed this deterministic approach following other previous studies pioneering the use of tumor forecasting in clinical settings (18, 21, 22, 28-31). Nevertheless, to maximize the utility of our predictive technology during AS, it would be necessary to enable tumor forecasting from the first imaging timepoint and then update the model as more mpMRI data becomes available. Indeed, the comparison of the results of the global calibration study (three mpMRI scans per patient) and those from the fitting-forecasting study (two imaging datasets per patient) suggests that incoming longitudinal data enables the refinement of the model parameterization and a more accurate representation of tumor morphology closer to the calibration horizon (e.g., lower RMSE or larger DSC, as observed herein). We believe that a probabilistic framework (18, 23) would enable a robust implementation of our predictive technology including the uncertainty quantification of input data and model outcomes (e.g., parameters, model-based biomarkers), the update of the model with new data collected during the course of AS (i.e., data assimilation), and tumor forecasting from the first imaging session (e.g., by sampling biomechanistic model parameters built in a separate population of patients). Since reproducibility and reliability are substantial barriers for the clinical adoption of new imaging biomarkers, uncertainty quantification could also contribute to the clinical translation of the model-based biomarkers proposed herein under variations in image acquisition as well as segmentation and registration errors. Fourth, our model is only informed by mpMRI data. Despite the wealth of spatiotemporal information provided by this imaging modality, PCa patients in AS are also monitored using PSA tests, which are often performed more frequently than mpMRI scans (2, 12, 13). Thus, combining the spatiotemporal model of tumor cell dynamics used here (i.e., Eqs. (2) or (3)) with an equation describing PSA dynamics could facilitate a more frequent update of the patient-specific parameters and include PSA dynamics in the construction of the PCa risk classifier (17, 32).

Finally, the simplicity of the model is a central limitation of this study. In particular, the biomechanistic model used herein tends to homogenize the intratumoral spatial distribution of the normalized tumor cell density map. This effect is caused by the Fisher-Kolmogorov equation, which we used to describe untreated PCa growth in our biomechanistic model (see Eqs. (2) and (3)). The natural solution of this equation is a travelling wavefront (51) that volumetrically expands a spatial region with non-zero tumor cell density to occupy other neighboring healthy regions, such that normalized tumor cell density progressively increases in the central part of the tumor and decreases along the travelling wavefront. Hence, although the initial conditions defined upon the normalized tumor cell density map extracted from the first mpMRI of each patient introduce information about the internal heterogeneity of the tumor architecture, the ensuing model simulation will progressively adapt the internal spatial distribution of normalized tumor cell density to match the natural solution of the Fisher-Kolmogorov equation. This effect contributes to explain the comparatively poorer performance of the personalized model at the third imaging timepoint with respect to the second one in both the global calibration and fitting-forecasting studies (see Figures 4 and 6), as well as the relatively lower local model-data agreement with respect to previous studies employing more advanced models (18, 23, 28, 29, 31). Additionally, the progressive model-driven spatial homogenization of the normalized tumor cell density maps does not enable to make accurate predictions of local GS values, for example, using the hyperbolic tangent mapping built to support the imaging preprocessing pipeline (see Figure 2). The analysis of these local GS maps would also require data on the spatial position of the PCa-positive biopsy cores and surgical specimen regions, but this information was not available for this study. Nevertheless, the prediction of local GS maps would be especially important to precisely identify the regions of the tumor for biopsy (2, 3, 16), to adapt the treating fields during radiotherapy (22, 23, 29), and, in general, to provide richer predictions of PCa progression during AS. Thus, future studies should investigate model extensions that incorporate mpMRI-informed spatial mechanisms that aim at retaining and evolving the intratumoral heterogeneity observed in imaging measurements of normalized tumor cell density maps. To achieve this goal, previous studies have leveraged heterogeneous and anisotropic parameterizations (18, 21, 23, 28, 29, 31), mechanical constraints to tumor cell mobility and proliferation (18, 21, 28, 29, 31), and information on tumor-supporting vascularity (18, 20, 21, 28, 59). In particular, given that the prostate is a confined organ in the pelvic region and that both a tumor and potentially coexisting benign prostatic hyperplasia may enlarge the prostate volume, the mechanical stress field within the organ can have a major impact on tumor dynamics (32). Thus, the extension of the model to include the mechanical effects on PCa growth constitutes a promising avenue of research.

In the future, we believe that further developing our computational modeling framework addressing the aforementioned limitations will enable the construction of a digital twin (27) for each individual patient enrolling in AS. These technologies not only enable a seamless integration of patient data and personalized model forecasts, but also exploit them to define specific metrics that aid the treating physician in making decisions about the timing and type of further testing or treatments, thereby providing a robust and systematic framework for the optimal management of each patient's tumor. By further extending our biomechanistic model to describe the response to standard first-line treatments for PCa after AS, the digital twin could also be trained to calculate therapeutic outcomes or expected survival, and, eventually, optimal regimens maximizing them while minimizing side-effects (18, 20-33). Future technological developments in mpMRI can also contribute to new and more accurate imaging measurements of PCa (3, 60), which can contribute to better inform our model and enhance the reliability of our model predictions. Thus, we posit that the work presented in this study constitutes an important first step toward the construction of a digital twin enabling a personalized, predictive paradigm for the clinical management of PCa.

## Conclusion

We have presented a clinical-computational framework for the personalized forecast of untreated, newly-diagnosed PCa growth during AS constructed upon a spatiotemporal biomechanistic model informed by patient-specific longitudinal mpMRI data. This technology tracks and predicts the three-dimensional development of the tumor within the patient's prostate geometry, while also enabling to anticipate the progression to higher risk-disease. Although further model development and investigation within larger cohorts are necessary, we posit that the personalized predictive technology presented herein can contribute to guide optimal clinical decision-making regarding the frequency and type of monitoring during AS, as well as the optimal time, modality, and planning of ensuing treatment.

## Acknowledgments

This project has received funding from the European Union's Horizon 2020 research and innovation programme under the Marie Skłodowska-Curie grant agreement No. 838786. J.S.H. and M.I.M. thank the National Institutes of Health for funding through R01EB027498. H.G. was partially funded by the National Science Foundation (CMMI 1852285). We also thank the Cancer Prevention and Research Institute of Texas (CPRIT) for support through CPRIT RR160005; T.E.Y is a CPRIT Scholar in Cancer Research. A.R. was partially supported by the MIUR-PRIN project XFAST-SIMS (No. 20173C478N). We thank the Texas Advanced Computing Center (TACC) for providing high-performance computational resources that contributed to the results presented in this work. The opinions, findings, and conclusions, or recommendations expressed are those of the authors and do not necessarily reflect the views of the funding agencies.

## Author contributions

Conceptualization: G.L., H.G., T.E.Y., T.J.R.H., A.R.

Data Curation: G.L., J.S.H., M.A.L.

Formal Analysis: G.L., J.S.H., M.I.M., H.G., T.E.Y., T.J.R.H., A.R.

Funding Acquisition: G.L., A.R.

Investigation: G.L., J.S.H., M.A.L., M.I.M., H.G., T.E.Y., T.J.R.H., A.R.

Methodology: G.L., J.S.H., M.A.L., M.I.M., H.G., T.E.Y., T.J.R.H., A.R.

Project Administration: G.L.

Resources: G.L., J.S.H., M.A.L.

Software: G.L., J.S.H.

Supervision: G.L.

Validation: G.L., J.S.H.

Visualization: G.L.

Writing - Original Draft Preparation: G.L., J.S.H., M.A.L.

Writing - Review & Editing: G.L., J.S.H., M.A.L., M.I.M., H.G., T.E.Y., T.J.R.H., A.R.

# Figures

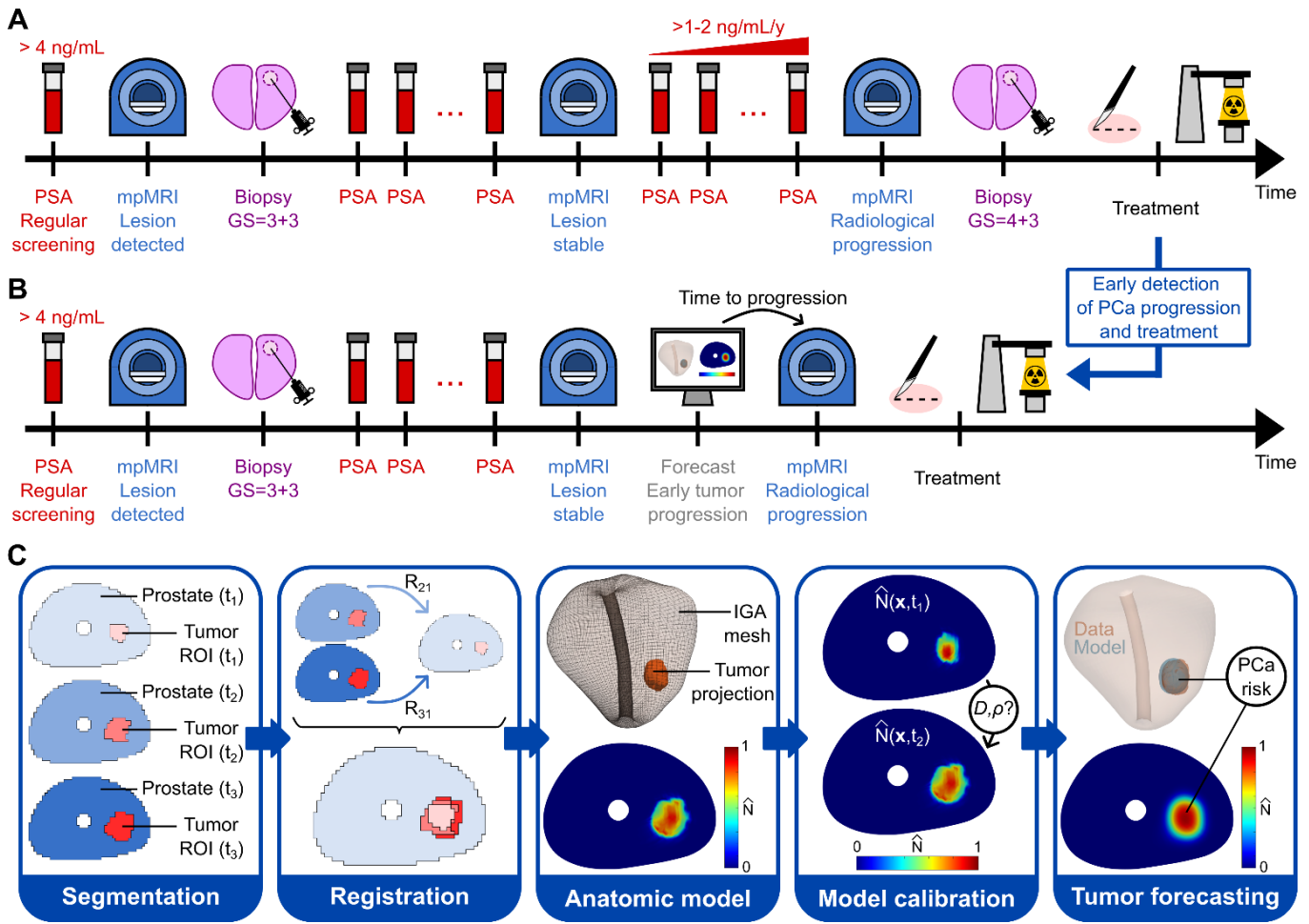

**Figure 1. Imaging-informed computational pipeline for tumor forecasting to support clinical decision-making in AS for PCa.** Panel A shows a standard-of-care AS protocol for an illustrative patient. Following the detection of a serum PSA level moderately larger than 4 ng/mL, the patient undergoes an initial mpMRI scan that finds an organ-confined cancerous lesion. This radiological lesion is then confirmed as PCa with GS=3+3 in an ensuing biopsy. Since the PCa risk is low, the patient enrolls in AS and periodic PSA tests are performed until the date of the next mpMRI scan in the AS protocol. This second imaging session does not reveal progression in the lesion, so the AS monitoring plan remains unchanged. However, the patient starts exhibiting a fast increase in PSA, which motivates an earlier imaging session before the originally prescribed date according to the AS protocol. This third mpMRI scan reveals radiological progression, which is further confirmed histopathologically as an upgrade to GS=4+3 in an ensuing biopsy. At this point, the patient is offered a radical treatment for PCa, which usually consists of surgery (i.e., radical prostatectomy) or radiotherapy (e.g., external beam radiotherapy, brachytherapy). Panel B illustrates the changes to the standard-of-care AS protocol after implementing the computational tumor forecasting pipeline presented in this study. The modified protocol is identical to the standard of care up to the second mpMRI. At this point, the longitudinal imaging data collected for the patient can be used to personalize our biomechanistic model of PCa growth and obtain a computational forecast of PCa growth over the patient's prostate anatomy. The forecast reveals progression toward high-risk PCa and provides the time up to this event. This prediction enables optimizing the timing of the third mpMRI to confirm progression and proceed to treatment. Thus,

our approach avoids PSA testing and biopsy after the second mpMRI, provides a personalized prediction of the patient's PCa progression that enables an early detection of this event, and supports the decision and optimal timing to perform treatment. Panel C summarizes the main steps in our computational pipeline for PCa forecasting during AS. In the cohort of this study ($n = 7$), all patients had 3 mpMRI scans. We first analyze the ability of our model to represent patient-specific PCa growth after being informed by 3 mpMRI scans. Then, we also investigate the ability of our model to forecast PCa growth when informed by only the first 2 mpMRI scans, and we use the third one to assess the predictive performance of the model. The first step of the computational pipeline is segmentation. We delineate the prostate and the tumor region of interest (ROI) on the longitudinal mpMRI data collected for the patient. After segmentation, the second and third mpMRI datasets and segments are co-registered with a non-rigid elastic method to the first one (registration transforms $R_{21}$ and $R_{31}$, respectively). Next, we build a virtual model of the prostate anatomy, consisting of a 3D isogeometric (IGA) mesh and we project the registered tumor ROIs onto it. Then, we map the ADC values within each tumor ROI to tumor cell density values, which are subsequently used to guide model calibration and determine the personalized model parameters. Finally, we perform a patient-specific tumor forecast, including the prediction of tumor volume, tumor cell density map, and PCa risk.

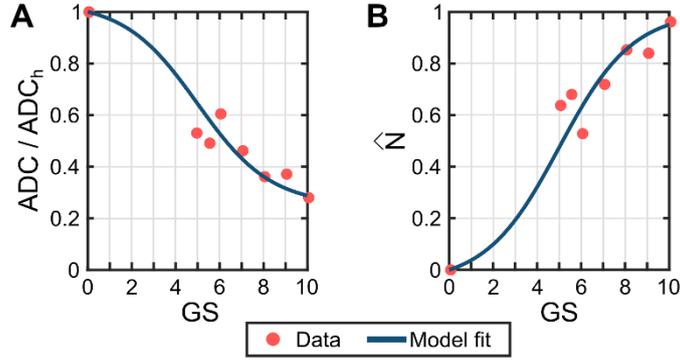

**Figure 2. Mapping ADC to N and GS.** Panel A shows a nonlinear least-squares fit of a hyperbolic tangent model (blue solid line) of the mean values of ADC in the tumor across GS groups in previous studies (red dots), which were normalized with respect to the mean ADC in healthy tissue ($ADC_h$) reported for each of them (41-45). We leverage a continuous, extended GS scale in which $0 < GS < 2$ indicates healthy and pretumoral tissue and $GS > 2$ corresponds to PCa (i.e., overlapping the standard discrete values of the GS used to assess histopathological samples of PCa; see Refs. (2, 4)). Further details on the hyperbolic tangent model and the fitting method are reported in Supplementary Methods S1. Panel B shows the mapping of the hyperbolic tangent model obtained for the normalized ADC values ($ADC/ADC_h$, panel A) to normalized tumor cell density values (i.e., $\hat{N}$, panel B). We use the linear mapping in Eq. (1), which introduces a negative proportionality between ADC and tumor cell density, as in other mpMRI-informed biomechanistic models of solid tumor growth (18, 21, 28, 29).

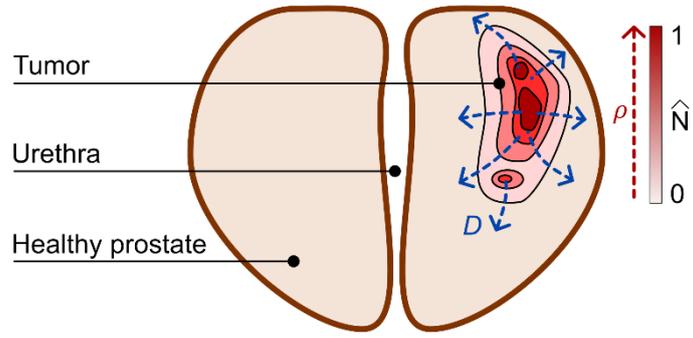

**Figure 3. Biomechanistic model of PCa growth during AS.** We describe the growth of newly-diagnosed untreated PCa in terms of the dynamics of the normalized tumor cell density, $\hat{N}(\boldsymbol{x},t)$, which may vary between $\hat{N} = 0$ far from the tumor to a maximum value of $\hat{N} = 1$ inside the tumor. The dynamics of the normalized tumor cell density is governed by the two mechanisms shown in this figure: (i) tumor cell mobility, which expands the tumor and is represented with a diffusion process controlled by parameter $D$ (i.e., the tumor cell mobility coefficient); and (ii) tumor cell net proliferation, which increases the local tumor cell density and is modeled as a logistic growth process controlled by parameter $\rho$ (i.e., tumor cell net proliferation rate). Since newly-diagnosed PCa cases eligible for AS are organ-confined, we further assume that tumor cells cannot leave the patient's prostate by enforcing $\nabla \hat{N} \cdot \boldsymbol{n} = 0$.

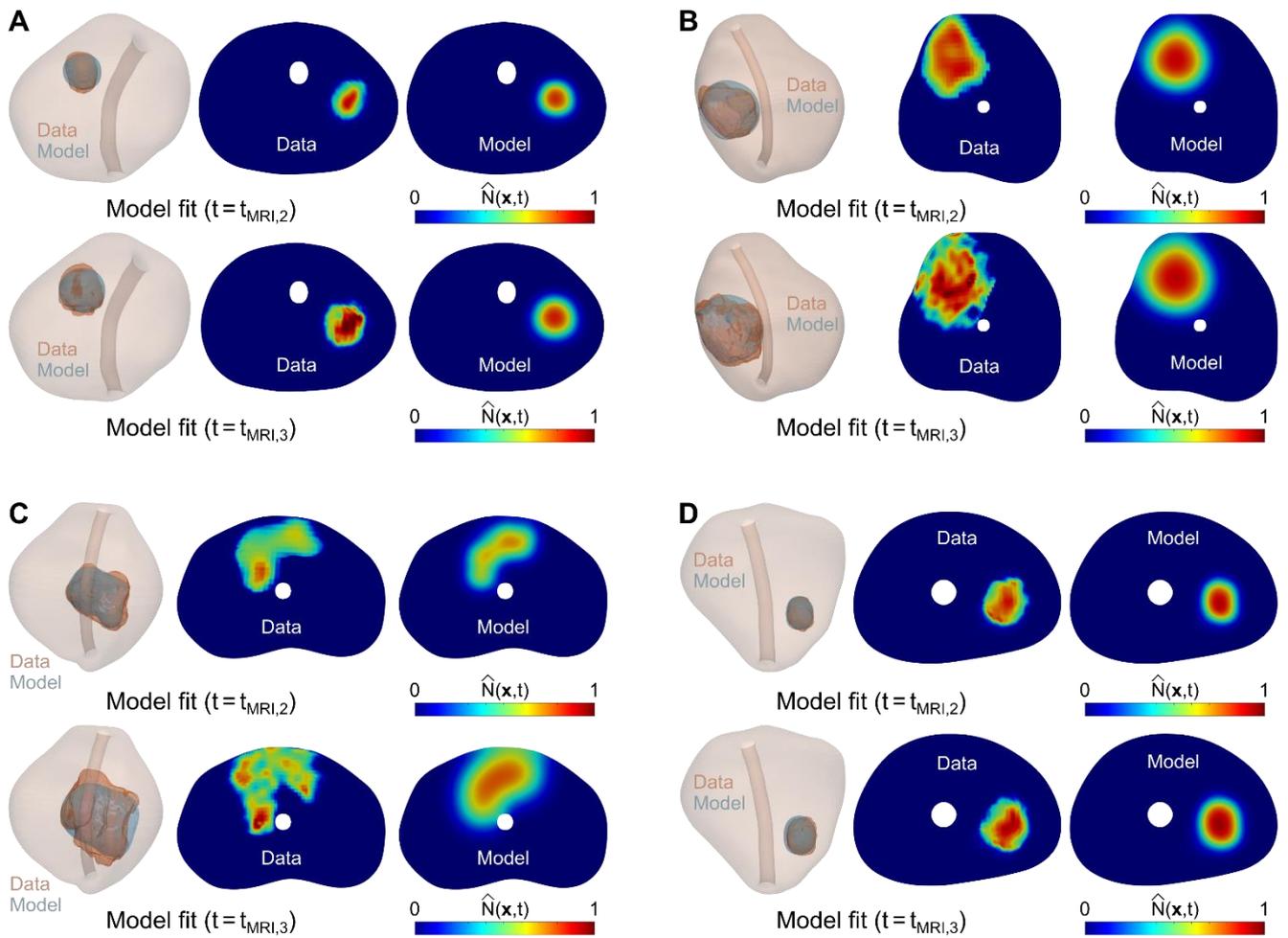

**Figure 4. Examples of personalized calculations of PCa growth in the global calibration scenario.** Panels A-D illustrate PCa growth at the dates of the 2$^{nd}$ mpMRI scan during model calibration ($t_{MRI,2}$; upper row in each panel) and the 3$^{rd}$ mpMRI scan ($t_{MRI,3}$; bottom row in each panel) in four patients. These results are illustrated using a 3D representation of the mpMRI-extracted prostate geometry of each patient including the imaging-measured and model-calculated tumor regions (red and blue volumes), along with an axial section of the prostate showing the normalized tumor cell density map obtained from the mpMRI data and using the personalized model (i.e., Eq. (3)).

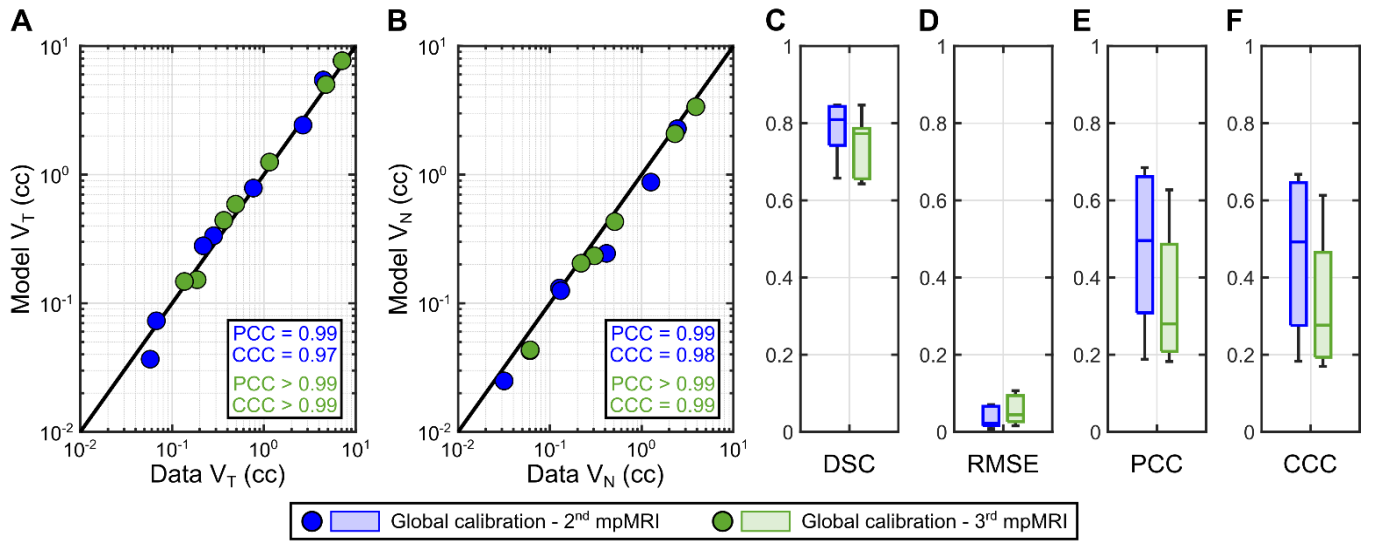

**Figure 5. Summary of metrics assessing the performance of the biomechanistic model during global calibration.** Panels A and B show unity plots to assess the agreement between the model estimation and mpMRI measurement of the tumor volume ($V_T$) and the total tumor cell volume ($V_N$) across the patient cohort ($n = 7$). These global tumor metrics were calculated at the dates of the 2nd and 3rd mpMRI scans (blue and green points). The unity plots in panels A and B also report the corresponding Pearson (PCC) and Concordance Correlation Coefficients (CCC) for the global tumor metrics at the dates of the 2nd and 3rd mpMRI scans. Panels C-F show the distributions of four local metrics assessing the agreement between the tumor cell density map calculated with our biomechanistic model and extracted from mpMRI data across the patient cohort in the global calibration scenario: Dice Similarity Coefficient (DSC), Root Mean Squared Error (RMSE), Pearson Correlation Coefficient (PCC), and Concordance Correlation Coefficient (CCC). These local metrics were calculated at the dates of the 2nd and 3rd mpMRI scans (blue and green boxplots).

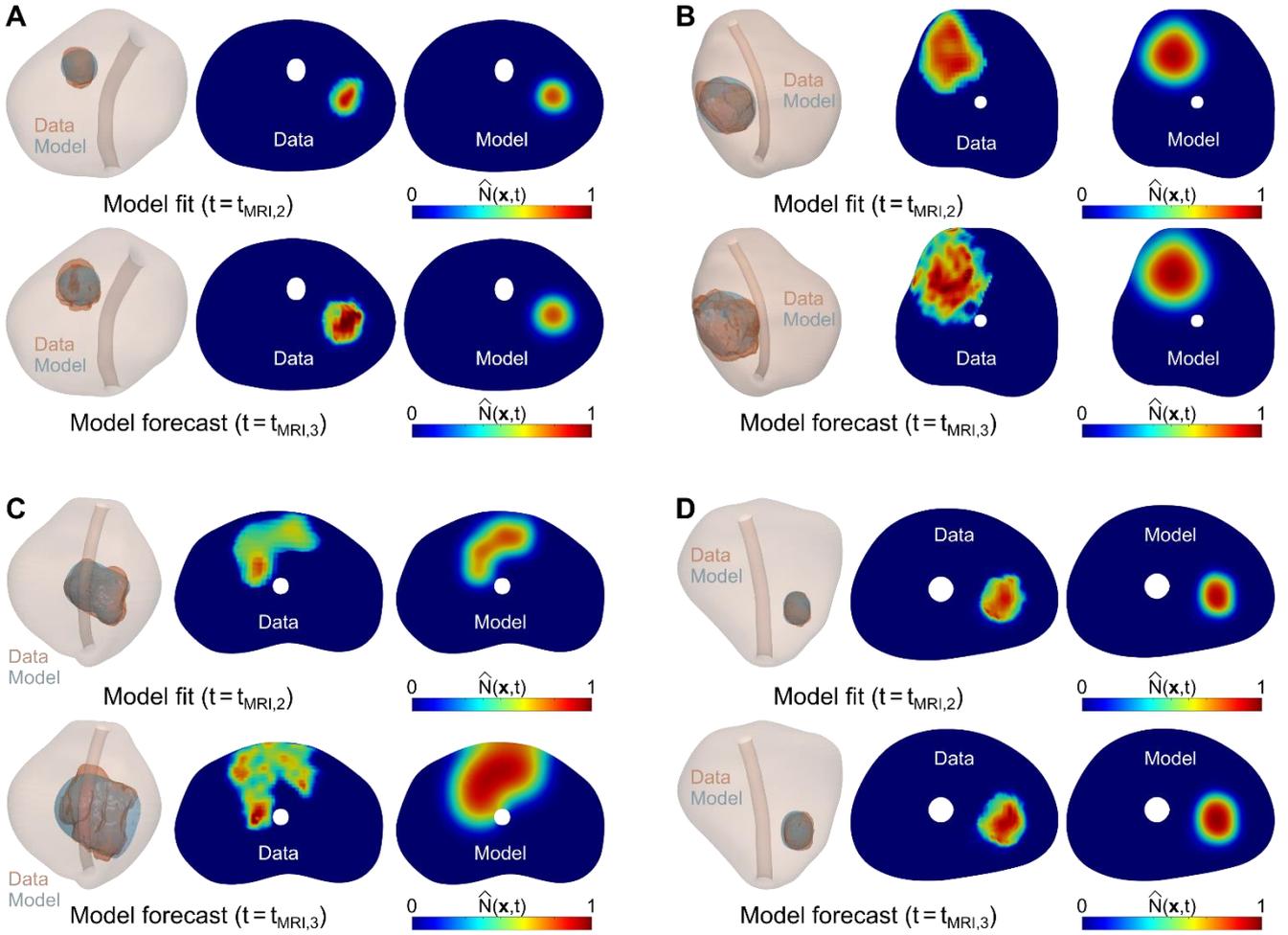

**Figure 6. Examples of personalized calculations of PCa growth in the fitting-forecasting scenario.** Panels A-D illustrate PCa growth at the date of the 2$^{nd}$ mpMRI scan during model calibration ($t_{MRI,2}$; upper row in each panel) and the ensuing prediction of tumor growth at the date of the 3$^{rd}$ mpMRI scan ($t_{MRI,3}$; bottom row in each panel) in four patients. The patients shown in this figure are the same considered in Figure 4. These results are illustrated using a 3D representation of the mpMRI-extracted prostate geometry of each patient including the imaging-measured and model-calculated tumor regions (red and blue volumes), along with an axial section of the prostate showing the normalized tumor cell density map obtained from the mpMRI data and using the personalized model (i.e., Eq. (3)).

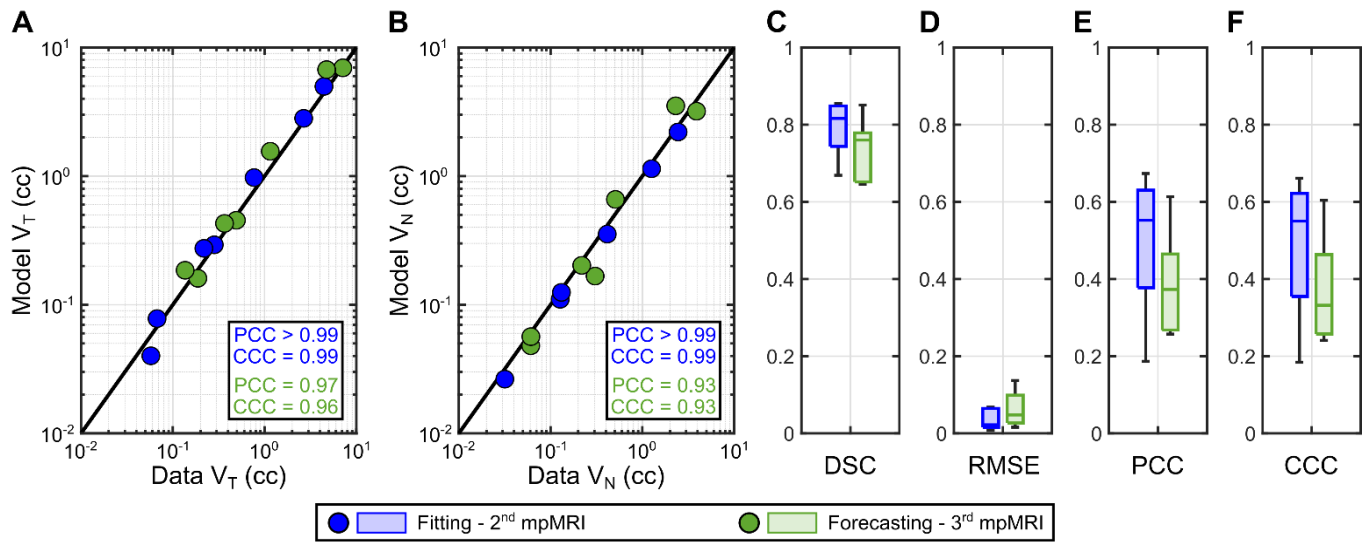

**Figure 7. Summary of metrics assessing the performance of the biomechanistic model in the fitting-forecasting scenario.** Panels A and B show unity plots to assess the agreement between the model estimation and mpMRI measurement of the tumor volume ($V_T$) and the total tumor cell volume ($V_N$) across the patient cohort ($n = 7$). These global tumor metrics were calculated at the dates of the 2nd and 3rd mpMRI scans (blue and green points). The unity plots in panels A and B further report the corresponding Pearson (PCC) and Concordance Correlation Coefficients (CCC) for the global tumor metrics at the dates of the 2nd and 3rd mpMRI scans. Panels C-F show the distributions of four local metrics assessing the agreement between the tumor cell density map calculated with our biomechanistic model and extracted from mpMRI data across the patient cohort in the global calibration scenario: Dice Similarity Coefficient (DSC), Root Mean Squared Error (RMSE), Pearson Correlation Coefficient (PCC), and Concordance Correlation Coefficient (CCC). These local metrics were calculated at the dates of the 2nd and 3rd mpMRI scans (blue and green boxplots).

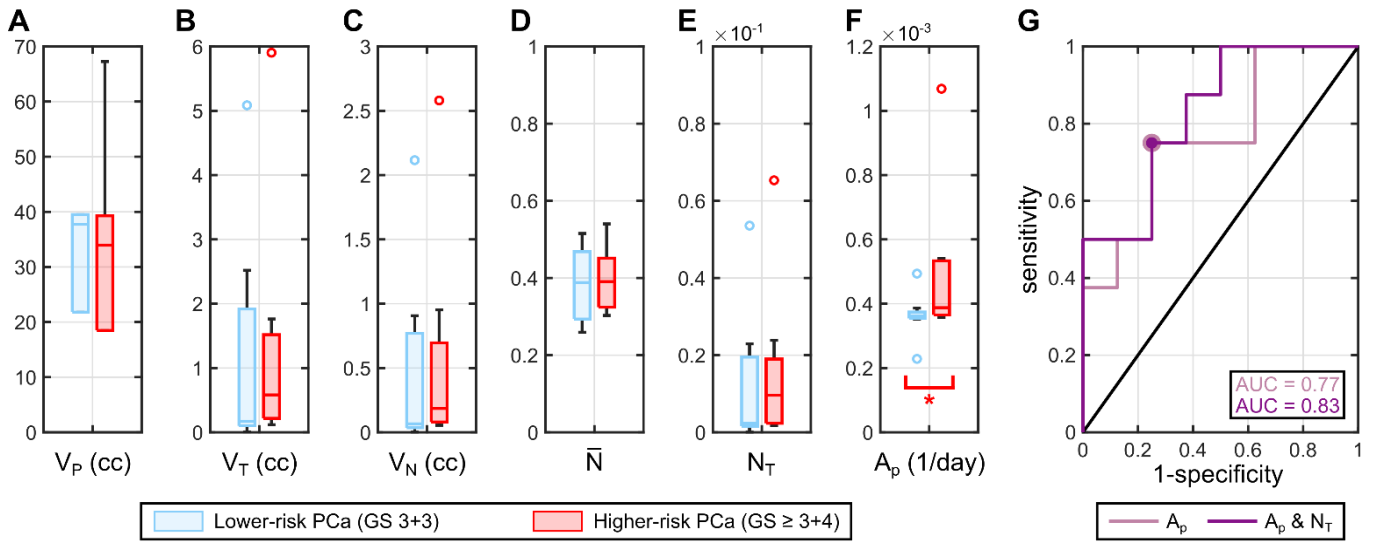

**Figure 8. Potential model-based biomarkers of higher-risk PCa**. Panels A-F show the distribution of six potential model-based biomarkers in lower-risk ($n = 8$) and higher-risk ($n = 8$) PCa, which were defined as tumors with GS=3+3 and GS≥3+4, respectively. These model-based biomarkers were calculated at the times were both a histopathological assessment and imaging measurement are available for each patient in the cohort ($n = 7$). In particular, the model-based markers in panels A-F are: prostate volume ($V_P$), tumor volume, total tumor cell volume ($V_N$), mean normalized tumor cell density ($\bar{N}$), total tumor index ($N_T$), and mean proliferation activity of the tumor ($A_p$). Outliers are represented as hollow circles and an asterisk indicates significance under a one-sided Wilcoxon rank-sum test ($p < 0.05$). Panel G shows the ROC curves for (i) the univariate logistic regression model constructed using the mean proliferation activity of the tumor (i.e., the only model-based marker that was significantly different between lower-risk PCa and higher-risk PCa), and (ii) the bivariate logistic regression model constructed using the mean proliferation activity of the tumor and the total tumor index (which is the combination of model-based markers that rendered the highest performance). The AUC of each ROC curve is reported within the plot, and the optimal performance point for both the univariate and bivariate logistic regression models operates at 75% sensitivity and specificity (bullet points on the ROC curves).

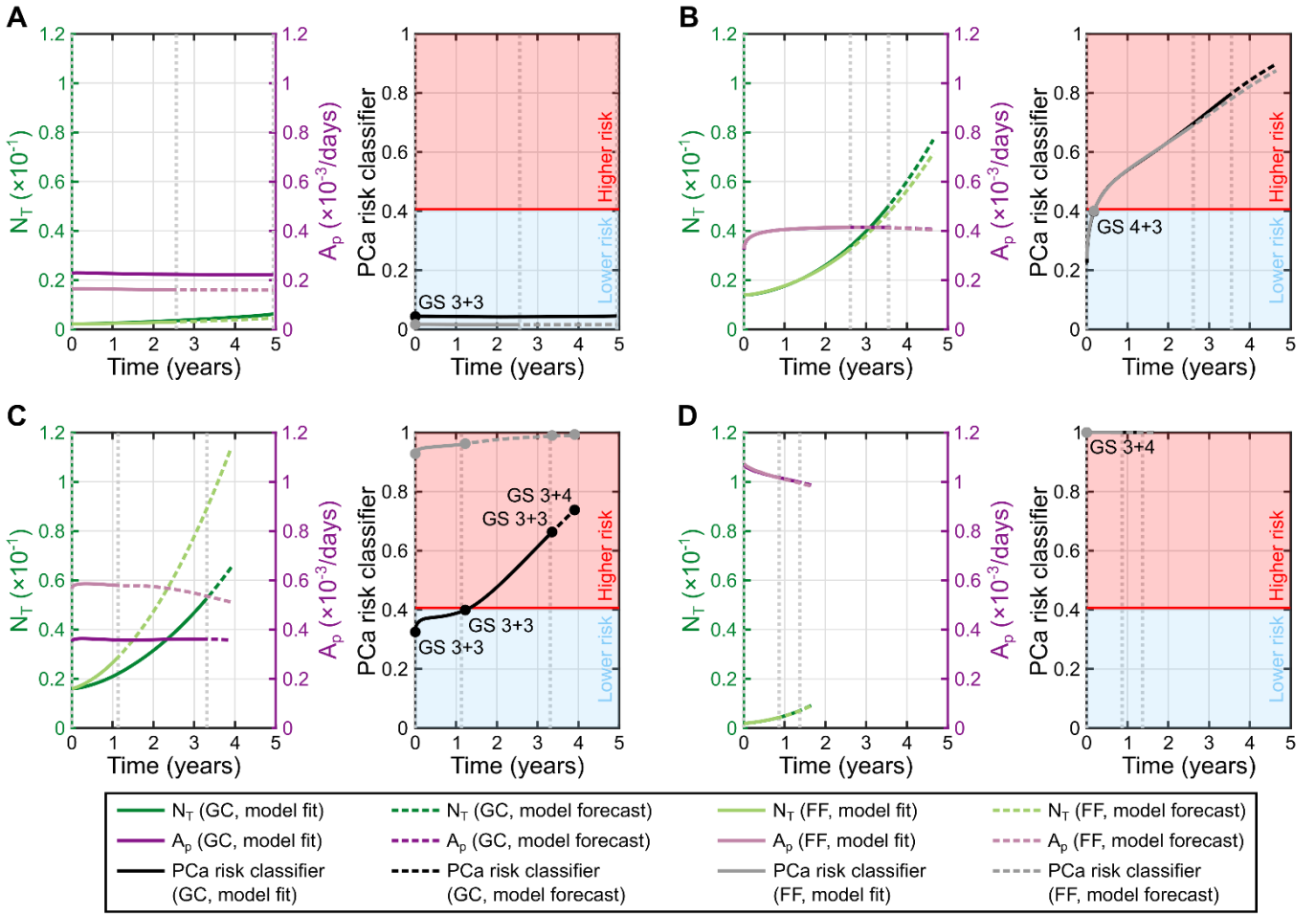

**Figure 9. Patient-specific forecasts of PCa risk**. Panels A-D show the time trajectories of the model-based markers involved in the calculation of our PCa risk classifier (left) as well as the trajectory of the latter (right) for four patients. The patients shown in this figure are the same considered in Figures 4 and 6. In all panels, darker hues represent results from the global calibration scenario (GC, see Figure 4), where the model is fit to the three mpMRI datasets from each patient, while lighter hues show results from the fitting-forecasting scenario (FF, see Figure 6), where the model is only fit to the first two mpMRI datasets from each patient. Dotted gray vertical lines in the background indicate the times of the mpMRI scans for each patient. Solid lines correspond to quantities calculated from the model fit, while dashed lines correspond to values calculated from model forecasts. The model-based markers of interest are the total tumor index ($N_T$, green curves, left vertical axis) and mean proliferation activity of the tumor ($A_p$, pink curves, right vertical axis). The PCa risk classifier was trained with the global calibration results (see Figure 8), yielding an optimal performance threshold that separates lower risk PCa (blue region) from higher-risk PCa (red region). The PCa risk at the times of histopathological assessment of the tumors (i.e., biopsy, surgery) is represented as a bullet point, and the corresponding GS values are annotated over the PCa risk trajectory from the global calibration scenario. In panel A, the patient exhibits a low-risk tumor during AS, which is correctly identified by our classifier in both computational scenarios. In panel B, our model consistently classifies the patient's tumor as higher-risk during the majority of AS in both computational scenarios. In panel C, the patient initially exhibits a lower-risk tumor that had progressed to higher-risk at the time of surgery (i.e., terminal GS value). In the fitting-forecasting scenario our model consistently predicts that the tumor is a higher-risk case, while, in the global calibration scenario, tumor progression is detected shortly after the second mpMRI scan. Importantly, in the fitting

forecasting scenario, our approach identifies a higher-risk tumor 1,011 days earlier than standard practice (i.e., assessment of the surgical specimen). In panel D, our model consistently identifies the tumor as a higher-risk case in both computational scenarios.

# Supplementary Information

# Patient-specific computational forecasting of prostate cancer growth during active surveillance using an imaging-informed biomechanistic model

G. Lorenzo[1,2], Jon S. Heiselman[3,4], Michael A. Liss[5], Michael I. Miga[3,6,7], Hector Gomez[8], Thomas E. Yankeelov[2,9,10], Alessandro Reali[1], Thomas J. R. Hughes[2]

[1]Department of Civil Engineering and Architecture, University of Pavia, Italy

[2]Oden Institute for Computational Engineering and Sciences, The University of Texas at Austin, USA

[3]Department of Biomedical Engineering, Vanderbilt University, USA

[4]Department of Surgery, Memorial Sloan-Kettering Cancer Center, USA

[5]Department of Urology, University of Texas Health San Antonio, USA

[6]Vanderbilt Institute for Surgery and Engineering, Vanderbilt University, USA

[7]Department of Neurological Surgery, Radiology, and Otolaryngology-Head and Neck Surgery, Vanderbilt University Medical Center, USA

[8]School of Mechanical Engineering, Weldon School of Biomedical Engineering, and Purdue Institute for Cancer Research, Purdue University, USA

[9]Livestrong Cancer Institutes and Departments of Biomedical Engineering, Diagnostic Medicine, and Oncology, The University of Texas at Austin, USA

[10]Department of Imaging Physics, The University of Texas MD Anderson Cancer Center, USA

**Corresponding author:**

Guillermo Lorenzo, PhD.

Department of Civil Engineering and Architecture

University of Pavia,

Via Ferrata 3, 27100, Pavia, Italia.

Email: guillermo.lorenzo@unipv.it, guillermo.lorenzo@utexas.edu

# Supplementary Methods

## S1. Hyperbolic tangent model fit for ADC ratio and continuous Gleason score

The apparent diffusion coefficient (ADC) measured via diffusion-weighted magnetic resonance imaging (DW-MRI) is known to decrease in prostate cancer (PCa) cases exhibiting higher Gleason score (GS), as reported in Refs. (41-45) in the main text. We exploited this observation for the semi-automatic segmentation of prostatic tumors after the manual delineation of a gross region of interest of the tumor lesion (see subsection on preprocessing of the imaging data in the Materials and Methods). Toward this end, we defined a quantitative relationship between ADC and GS, such that we could automatically identify the subregion of the gross tumor delineation with values of ADC representing mpMRI-observable tumors, which usually exhibit GS ≥ 3+3. To facilitate the application of our methods to ADC maps obtained for each patient at different time points and across patients, we leveraged the ADC ratio ($ADC_r = ADC/ADC_h$, where $ADC_h$ is a representative measurement of ADC in each patient's healthy prostatic tissue), which has also been shown to exhibit a decreasing trend with increasing Gleason score (41-45). Furthermore, we extended the discrete definition of GS used in the clinic, whereby this metric can only take natural numbers in the range [2, 10], to a continuous definition in the range [0, 10]. Hence, the values in the interval [0, 2] are assumed to represent healthy and pre-tumoral lesions. The rationale for the continuous definition was to account for the changing ADC as the tumor evolves toward higher GS, such that these continuous dynamics could then be matched to the spatiotemporal evolution of the normalized tumor cell density maps calculated with our biomechanistic model of PCa growth. To capture the decrease in the ADC ratio with increasing GS, we adopted a hyperbolic tangent function because it provided two horizontal asymptotes that can be used to account for two additional features of interest: (i) a plateauing trend in the decrease of ADC ratio for the highest GS values, which aligns with a higher tumor cell density approaching tissue carrying capacity in those tumors (see Refs. (46-48) in the main text); and (ii) a plateau for values of ADC ratio close to zero (i.e., healthy tissue) followed by a slow decrease of the ADC ratio in pre-tumoral lesions and low GS tumors, which makes them practically indistinguishable from healthy tissue and matches the inability of mpMRI to accurately identify PCa cases with GS < 3+3.

The mathematical formulation of the hyperbolic tangent function that we used in this work is given by

$$ADC_r = a - b\tanh(c(GS - d)), \tag{S1}$$

where $a$, $b$, $c$, and $d$ are empirical fitting constants. We enforced $ADC_r = 1$ for $GS = 0$, such that the ADC in healthy tissue satisfied $ADC = ADC_h$. By introducing this constraint in Eq. (S1), we could eliminate parameter $a$ and simplify the fitting problem to the determination of the other empirical constants using the following equation:

$$ADC_r = 1 + b(\tanh(-cd) - \tanh(c(GS - d))). \tag{S2}$$

We calculated the values of the empirical constants $b$, $c$, and $d$ that provided the best fit of Eq. (S2) to the average of the ADC ratios obtained from the mean $ADC_h$ and mean GS-specific ADC values reported in the five studies mentioned above (i.e., Refs. (41-45) in the main text). Curve fitting was carried out using the *fit* function from the Curve Fitting Toolbox in MATLAB (R2021b; The Mathworks, Natick, MA). The maximum number of iterations and model evaluations were both set at 10,000. The admissible range of values for the empirical constants were $b \in [0, 1]$, $c \in [0.30, 0.50]$, and $d \in [5, 7]$. The initial guess of these empirical constants was set at the middle point of their corresponding admissible value range. Hence, curve fitting resulted in the following definition of Eq. (S1):

$$ADC_r = 0.64 - 0.39\tanh(0.3(GS - 5)). \tag{S3}$$

The lower asymptote of the hyperbolic tangent function in Eq. (S3) can be calculated as the limit of $ADC_r$ when $GS \to \infty$, and was used to define $ADC_{min} \approx 0.25 ADC_h$. This quantity represents the minimum admissible value of the ADC ratio in PCa, and it enabled the conversion of ADC map measured over a patient's tumor to a normalized tumor cell density map ($\widehat{N}(x, t)$) via the linear mapping given by

$$\widehat{N}(x, t) = \frac{ADC_h - ADC(x, t)}{ADC_h - ADC_{min}}. \tag{S4}$$

Importantly, Eq. (S4) accounts for the decrease in ADC with increasing tumor cell density in higher GS tumors (see Refs. (41-48) in the main text). Furthermore, Eq. (S4) has been used in the preprocessing step of the computational pipeline of several successful tumor forecasting studies employing mpMRI-informed biomechanistic models (see Refs. (18, 21, 28, 29) in the main text).

# Supplementary Figures

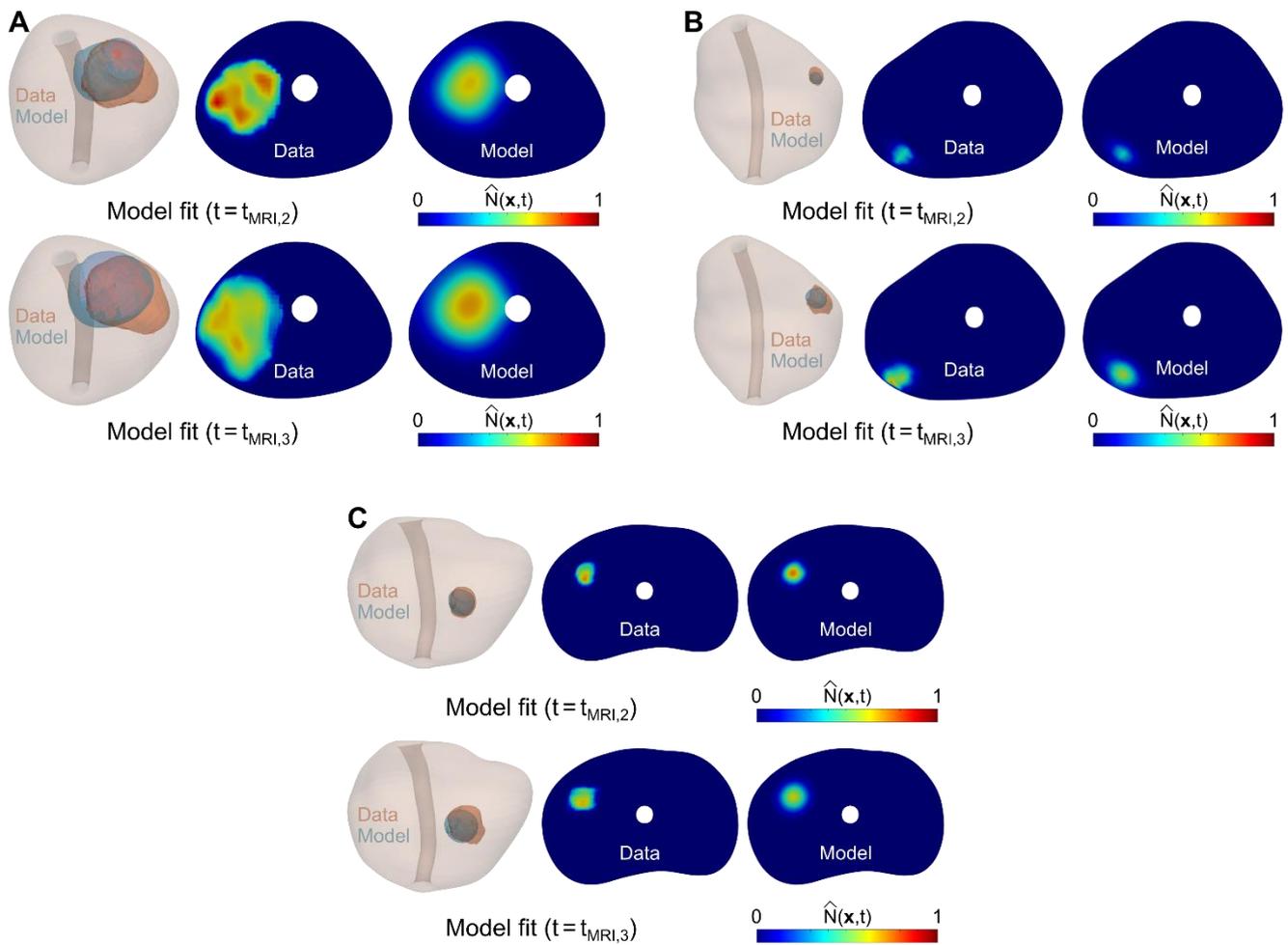

**Supplementary Figure S1. Examples of personalized calculations of PCa growth in the global calibration scenario.** Panels A-C illustrate PCa growth at the dates of the 2nd mpMRI scan during model calibration ($t_{MRI,2}$; upper row in each panel) and the 3rd mpMRI scan ($t_{MRI,3}$; bottom row in each panel) for the remaining three patients not shown in Figure 4 of the main text. These results are illustrated using a 3D representation of the mpMRI-extracted prostate geometry of each patient including the imaging-measured and model-calculated tumor regions (red and blue volumes), along with an axial section of the prostate showing the normalized tumor cell density map obtained from the mpMRI data and using the personalized model (i.e., Eq. (3) in the main text).

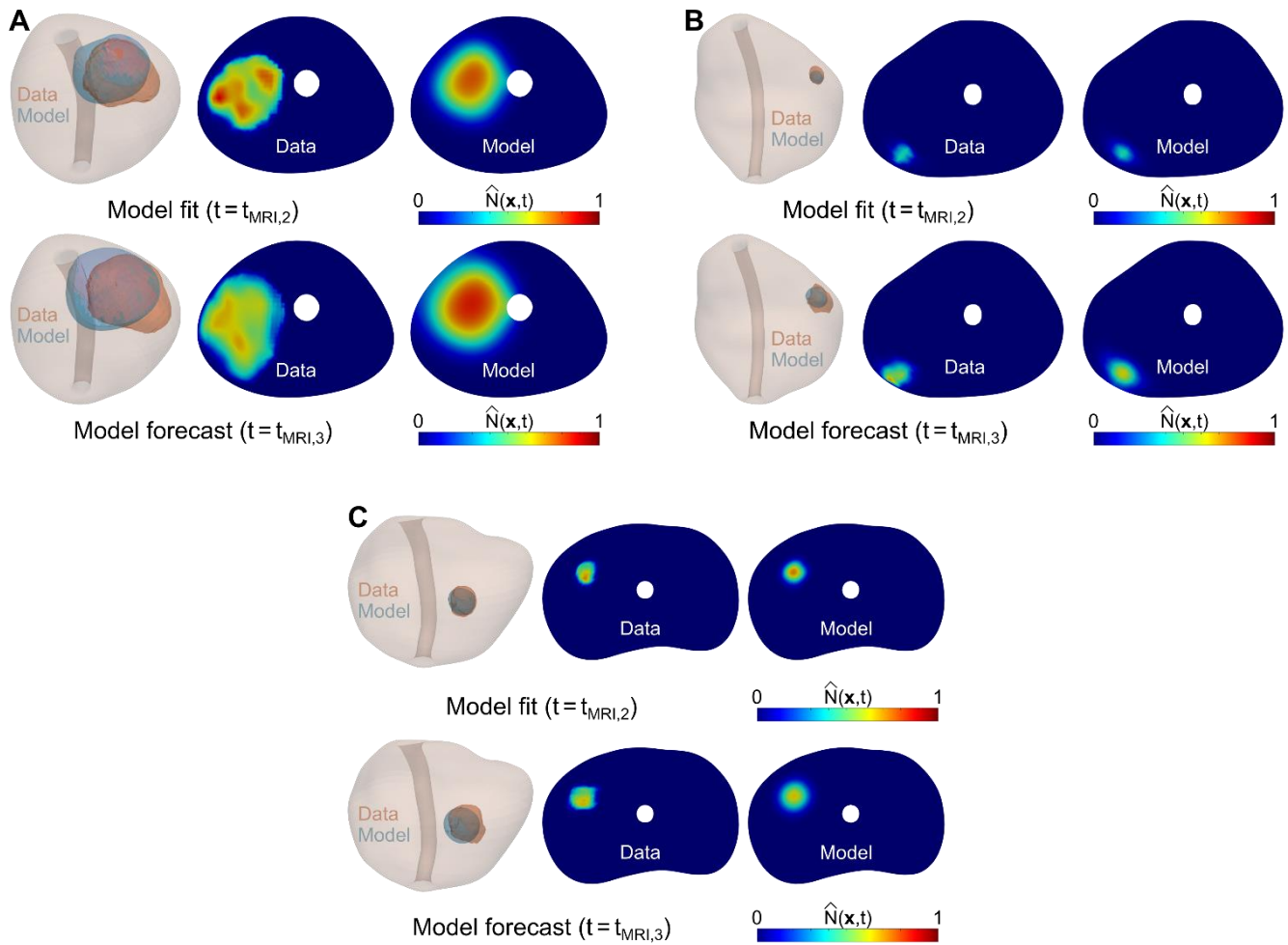

**Supplementary Figure S2. Examples of personalized calculations of PCa growth in the fitting-forecasting scenario.** Panels A-C illustrate PCa growth at the date of the 2$^{nd}$ mpMRI scan during model calibration ($t_{MRI,2}$; upper row in each panel) and the ensuing prediction of tumor growth at the date of the 3$^{rd}$ mpMRI scan ($t_{MRI,3}$; bottom row in each panel) for the remaining three patients not shown in Figure 6 of the main text. The patients shown in this figure are the same considered in Supplementary Figure S1. These results are illustrated using a 3D representation of the mpMRI-extracted prostate geometry of each patient including the imaging-measured and model-calculated tumor regions (red and blue volumes), along with an axial section of the prostate showing the normalized tumor cell density map obtained from the mpMRI data and using the personalized model (i.e., Eq. (3) in the main text).

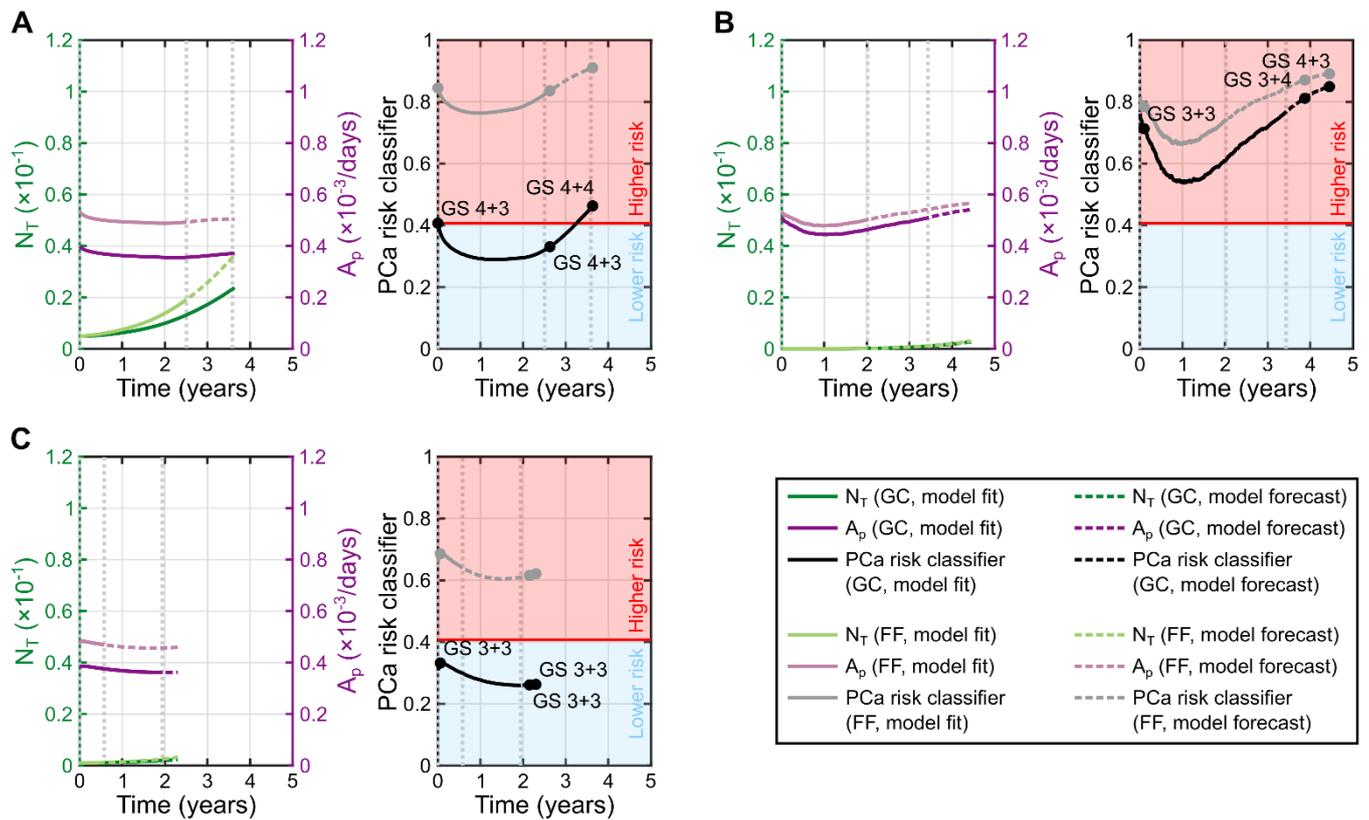

**Supplementary Figure S3. Patient-specific forecasts of PCa risk.** Panels A-C show the time trajectories of the model-based markers involved in the calculation of our PCa risk classifier (left) as well as the trajectory of the latter (right) for the remaining three patients not shown in Figure 9 of the main text, respectively. The patients shown in this figure are the same considered in Supplementary Figures S1 and S2. In all panels, darker hues represent results from the global calibration scenario (GC, see Supplementary Figure S1), where the model is fit to the three mpMRI datasets from each patient, while lighter hues show results from the fitting-forecasting scenario (FF, see Supplementary Figure S2), where the model is only fit to the first two mpMRI datasets from each patient. Dotted gray vertical lines in the background indicate the times of the mpMRI scans for each patient. Additionally, solid lines correspond to quantities calculated from the model fit, while dashed lines correspond to values calculated from model forecasts. The model-based markers of interest are the total tumor index ($N_T$, green curves, left vertical axis) and mean proliferation activity of the tumor ($A_p$, pink curves, right vertical axis). The PCa risk classifier was trained with the global calibration results (see Figure 8 in the main text), yielding an optimal performance threshold that separates lower risk PCa (blue region) from higher-risk PCa (red region). The PCa risk at the times of histopathological assessment of the patients' tumors (i.e., biopsy, surgery) is represented as a bullet point, and the corresponding GS values are annotated over the PCa risk trajectory from the global calibration scenario. In panel A, in the fitting-forecasting scenario the PCa risk classifier consistently identifies the tumor as a higher-risk case and anticipates tumor progression from GS 4+3 to GS 4+4 using the personalized predictions at the second imaging timepoint, which is 410 days (i.e., ~1.1 years) earlier than the final histopathological assessment at surgery. The global calibration scenario confirms higher-risk disease at the final histopathological assessment, but suggests a lower risk at the times of the first two mpMRI scans and biopsies. In panel B, the tumor is consistently classified as a higher-risk case in both computational scenarios. In this case, the PCa forecasts after the second mpMRI scan in the fitting-forecasting scenario enable the detection of tumor progression 677 days (i.e., ~1.9 years) before surgery. The tumor in panel C is consistently classified as a higher-risk case using the predictions from the fitting-forecasting study, but assimilation of the third mpMRI dataset in the global calibration scenario correctly classified the tumor as a lower-risk case.